\documentclass[twocolumn]{aastex631}

\usepackage{amsmath, amssymb}
\usepackage{savesym}
\savesymbol{tablenum}
\usepackage{siunitx}
\restoresymbol{SIX}{tablenum}
\usepackage{xspace}
\usepackage[super]{nth} 
\usepackage{multirow}
\usepackage[T1]{fontenc}
\usepackage{longtable}
\usepackage{rotating,tabularx}

\newcommand{\kepler}{\emph{Kepler}\xspace}
\newcommand{\gaia}{\emph{Gaia}\xspace}
\defcitealias{DressingCharbonneau2015}{DC15}
\usepackage[normalem]{ulem}

\received{April 7, 2025}
\revised{November 12, 2025}
\accepted{December 15, 2025}

\begin{document}

\title{Modeling the Impact of Unresolved Stellar Companions on Detection Sensitivity in Kepler’s Small Planet Occurrence Rates}

\correspondingauthor{Galen J. Bergsten}
\email{gbergsten@stsci.edu}

\author[0000-0003-4500-8850]{Galen J. Bergsten}
\affil{Lunar and Planetary Laboratory, The University of Arizona, Tucson, AZ 85721, USA}
\affil{NASA Exoplanet Science Institute-Caltech/IPAC, Pasadena, CA 91125, USA}
\affil{Space Telescope Science Institute, Baltimore, MD 21218, USA}

\author[0000-0002-5741-3047]{David R. Ciardi}
\affil{NASA Exoplanet Science Institute-Caltech/IPAC, Pasadena, CA 91125, USA}

\author[0000-0002-8035-4778]{Jessie L. Christiansen}
\affil{NASA Exoplanet Science Institute-Caltech/IPAC, Pasadena, CA 91125, USA}

\author[0000-0002-2361-5812]{Catherine A. Clark}
\affil{NASA Exoplanet Science Institute-Caltech/IPAC, Pasadena, CA 91125, USA}

\author[0000-0001-7962-1683]{Ilaria Pascucci}
\affil{Lunar and Planetary Laboratory, The University of Arizona, Tucson, AZ 85721, USA}

\author[0000-0001-8189-0233]{Courtney D. Dressing}
\affiliation{Department of Astronomy, University of California, Berkeley, Berkeley, CA 94720, USA}

\author[0000-0003-3702-0382]{Kevin K.\ Hardegree-Ullman}
\affil{NASA Exoplanet Science Institute-Caltech/IPAC, Pasadena, CA 91125, USA}

\author[0000-0003-2527-1598]{Michael B. Lund}
\affil{NASA Exoplanet Science Institute-Caltech/IPAC, Pasadena, CA 91125, USA}

\newcommand {\galen}[1]  {\textit{\textcolor{magenta}{[Galen: #1]}}}

\begin{abstract}
Unresolved stellar companions can cause both under-estimations in the radii of transiting planets and over-estimations of their detectability, affecting our ability to reliably measure planet occurrence rates. To quantify the latter, we identified a control sample of 198 \kepler{} stars with sensitivity to Earth-like planets if they were single stars, and imaged them with adaptive optics. In 20\% of systems, we detected stellar companions that were close enough to go unresolved in \kepler{} observations. We calculated the distribution of planet radius correction factors needed to adjust for these observed companions, along with simulations of undetected companions to which our observations were not sensitive. We then used these correction factors to optimize an occurrence rate model for small close-in planets while correcting \kepler{}'s detection efficiency for the presence of unresolved companions, and quantified how this correction affects occurrence estimates. Median occurrence rates for small planets between 2--100\,days increased by an average factor of $1.08-1.19$ (depending on statistical treatments), with the largest differences found for smaller planets at larger orbital periods. We found that the frequency of Earth-sized planets in the habitable zone ($\eta_\oplus$) increased by a factor of ${1.18}_{-0.66}^{+0.43}$ -- ${1.46}_{-0.83}^{+0.53}$ when accounting for the effect of unresolved companions on \kepler{}'s detection sensitivity.
\end{abstract}


\section{Introduction} \label{sec:intro}
Approximately half of Solar-type stars in the local neighborhood have bound stellar companions \citep[][and references therein]{Offner2023}. The presence (or absence) of stellar companions can have a significant influence on our ability to detect and correctly characterize transiting exoplanets. First, the additional star adds ``third light'' to the system which can dilute a transit signal as observed by transit surveys such as \kepler, K2, or TESS. The shallower transit leads to an underestimate of the planet's radius, i.e., planets appear smaller than they actually are \citep{Ciardi2015}. Second, transiting planets in systems with multiple stars have a probability of orbiting the secondary star (and not the primary), and thus the derived planet radius can be significantly larger than initially estimated by assuming the planet orbits a single star \citep[e.g.,][]{Hirsch2017AJ}. Third, the companion star may be of sufficient brightness to fill in the transit completely, preventing the detection of the planet altogether \citep[e.g.,][]{Lester2021}. These effects combine to influence the derivation of the planet radii and associated occurrence rates. When planetary occurrence rates -- particularly those of Earth-sized planets -- have been derived from transit surveys, it is typical to exclude obvious multi-star systems from the input stellar catalog and assume all remaining targets are single \citep[e.g.,][]{Bryson2021, Zink2021, Bergsten2022, Gan2023}. But if the frequency of Earth-sized planets in the habitable zone ($\eta_\oplus$) is to be truly understood, then stellar companions must be accounted for \citep[e.g.,][]{Hirsch2017AJ, Bouma2018}. 

High-resolution imaging has been used to search for close-in (bound and unbound) stellar companions as part of the standard process for transiting planet validation and confirmation. As a result, a significant fraction of the \kepler candidates have been imaged using near-infrared adaptive optics and optical speckle imaging (see e.g., \citealp{Furlan2017}, which also compiles work from \citealp{Adams2012, Adams2013, Baranec2016, Cartier2015, Dressing2014, Everett2012, Everett2015, Gilliland2015, Horch2012, Horch2014, Howell2011, Kraus2016, Law2014, Lillo-Box2012, Lillo-Box2014, Wang2015a, Wang2015b, Wang2015c, Ziegler2017}). However, the construction of the \kepler occurrence rate products (e.g., the \kepler candidate list, the completeness, and the reliability products) do not take into account the blending of nearby companions that do not appear in the (seeing-limited) \kepler Input Catalog \citep{Brown2011,Mullally2015}. As a result, occurrence rate estimates based on the \kepler products do not take into account the effects of stellar companions on either our ability to detect transiting planets or to characterize the size of the planets accurately.

A first attempt at understanding the corrections needed to address unresolved companions was made with a sample of \kepler targets observed for multiplicity with the Shane AO system \citep{savel2020} where they found the occurrence of Earth-like planets to increase by a relative factor of 1.06, with a factor of 1.26 increase for super-earths and sub-Neptunes.  That survey, however, was relatively small (71 stars), was observed at Lick with the 3m, and was published prior to the release of \gaia DR3 \citep{GaiaDR3}.  

In this work, we aim to better assess the effect that unresolved stellar companions may have on the derivation of \kepler{} small planet occurrence rates via misestimations of detection sensitivity. We first identified a larger control sample of \kepler{} stars that, if single, would have been amenable to detecting Earth-sized planets at Earth-like insolations. We surveyed these stars with high-resolution adaptive optics imaging using the 5\,m Hale telescope at Palomar Observatory, and assessed their potential binarity from \gaia DR3. We then used measurements of the companions in this control sample and the occurrence rate infrastructure from \citet{Bergsten2022} to directly compare the derived occurrence rates when the stars are assumed to be single versus when the influence of stellar multiplicity is included. By doing this comparison using the same occurrence rate model, we derived a relative probabilistic correction as a function of planet size and orbital separation. We then evaluated how these corrections impact habitable zone occurrence estimates, in order to quantify how statistically accounting for unresolved stellar companions could influence predictions for Earth-like planets.

The observation and characterization of the control sample is described in Section~\ref{sec:sample}, and the occurrence rate modeling methodology is described in Section~\ref{sec:occ}. In Section~\ref{sec:results}, we present a comparison of planetary occurrence rates in our models with/without companions, and summarize our findings in Section~\ref{sec:end}.

\section{Constructing the Control Sample}\label{sec:sample}

\subsection{Observations}

\begin{figure}
    \centering
    \includegraphics[width=\linewidth]{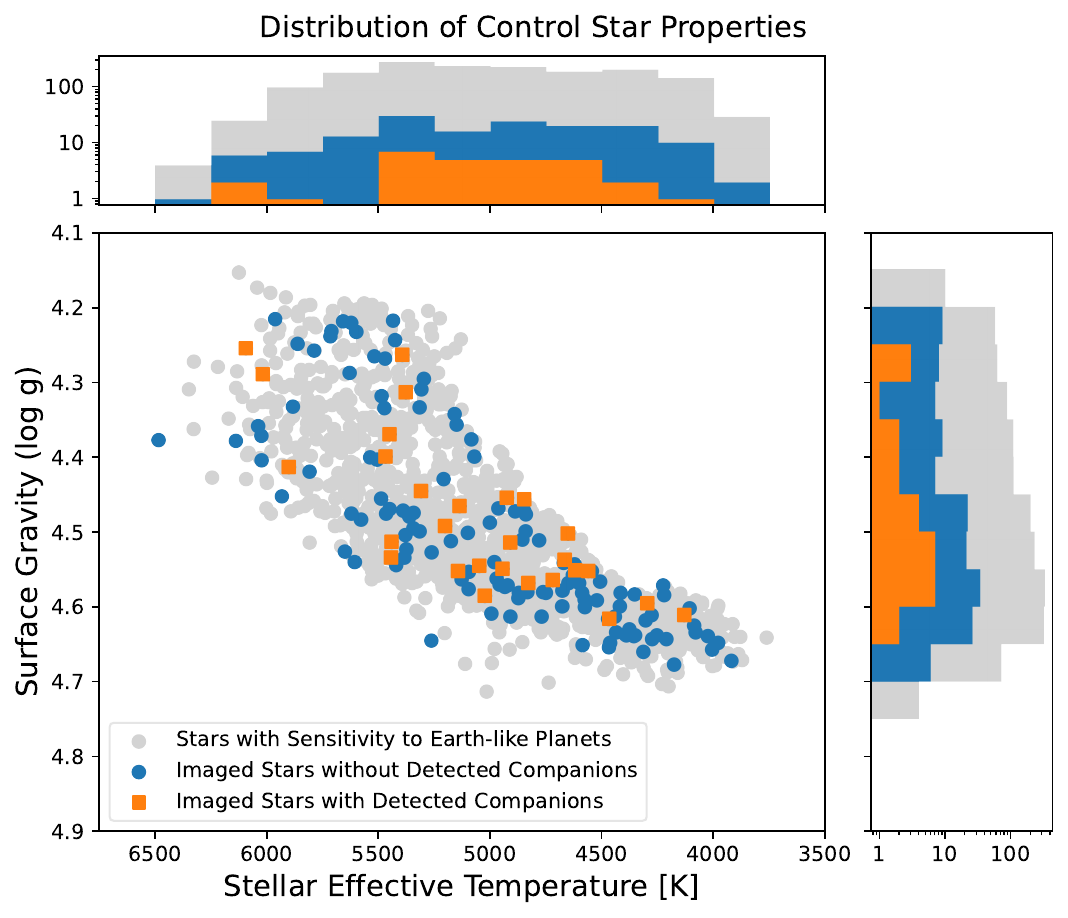}
    \caption{The sample of \kepler{} FGK dwarf stars around which a 1 R$_\oplus$ planet receiving the same insolation as Earth could have been detected. The full set of 2,136 stars are shown in gray circles, using the stellar effective temperatures and (log) surface gravities available in \citet{Berger2020}. The randomly selected subset of 198 Sun-like control stars we observed are shown in color: the set of control stars for which we found companions are shown in orange squares, while the control stars we did not detect companions around are shown in blue circles. Side panels show the corresponding histograms for the \textbf{top:} stellar effective temperature and \textbf{right:} surface gravity distributions.}
    \label{fig:control}
\end{figure}

From the \kepler DR25 stellar parameters \citep{Mathur2017} and measured noise properties \citep{Thompson2018}, there are 2,136 \kepler dwarf targets for which the stellar radius, combined differential photometric precision \cite[CDPP;][]{Christiansen2012,Christiansen2020}, and observing baseline are amenable to the detection of 1~$R_{\oplus}$ planets receiving the same insolation as the Earth with the \kepler pipeline. From this set, we randomly selected a control sample of 198 Sun-like (FGK, with stellar masses between $0.56 - 1.63$\,M$_\odot$; \citealp{PecautMamajek}) targets that encompassed the range of stellar effective temperatures, surface gravities, and magnitudes represented by the full set \citep[see also][]{savel2020}. To verify the control sample is appropriately representative of the full set, we compared their effective temperature and surface gravity distributions through two-sample Kolmogorov-Smirnov tests. We found p-values of 0.93 and 0.38 respectively, affirming that the two populations are not significantly different. The distribution of stellar properties for the control sample and full set of 2,136 stars are included in Figure~\ref{fig:control}.

Each of the 198 \kepler field stars were observed between 2017 and 2019 with the near-infrared adaptive optics system on the 5\,m Hale telescope at Palomar Observatory. The observations at Palomar were obtained with the near-infrared camera PHARO \citep{hayward2001} which sits behind the P3K natural guide AO system \citep{dekany2013}. PHARO has a pixel scale of 0.025\arcsec\ per pixel with a field of view of 25.6\arcsec. Targets were observed with a mix of \SI{2}{\micro\meter} filters ($K$, $K_s$, $Br_\gamma$) -- the choice of filter was made at the time of observations depending on the brightness of the target and conditions. The $Br_\gamma$ filter was typically used on the brightest targets and best nights, as the narrow bandwidth yields a better AO-correction than the wider $K$ or $K_s$ filters. If a stellar companion was immediately obvious, a second observation was taken in either $J$ or $H_{cont}$ (PHARO does not have a $J_{cont}$ filter). The $J$-filter was chosen primarily to yield $J-K$ color which is a stronger ``lever-arm'' than $H-K$ for a classification of the companions. However, conditions did not always allow for good $J$-band corrections with the AO-system, in which case $H_{cont}$ was obtained instead. The integration time for each target was set to ensure that the peak of the primary target remained in the linear range of the detectors.

\begin{figure}
    \centering
    \includegraphics[width=\linewidth]{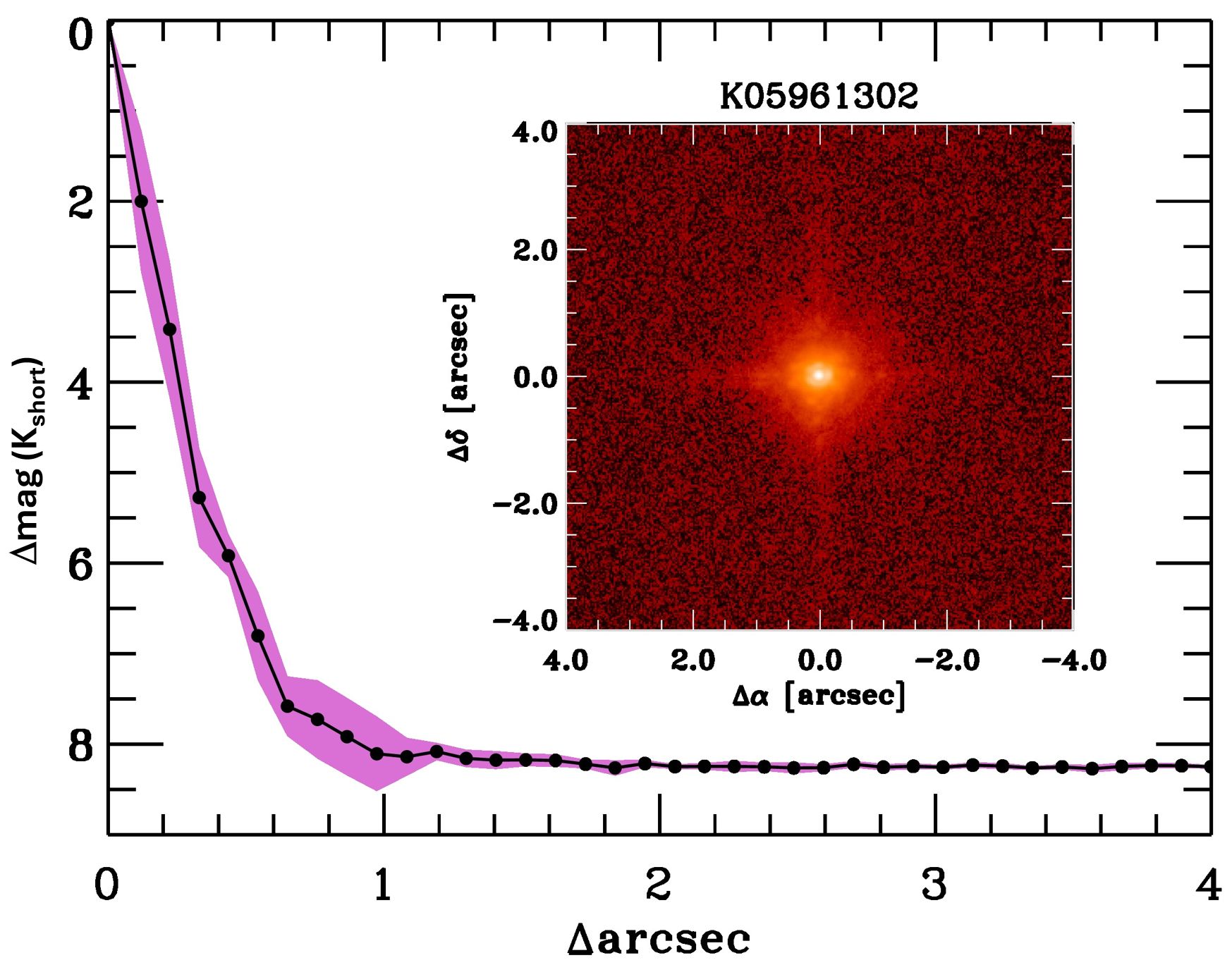}
    \caption{Example AO adaptive image and sensitivity curve for KIC 5961302 -- a V=13.7mag K4V \kepler target in our control sample. The resolution of the image is 0.109\arcsec (FWHM of the PSF).  The black points represent the measured sensitivity relative to the brightness of the target in FWHM radial steps from the target.  The purple represents the azimuthal rms dispersion of the sensitivity measurements. The inset image is a 4\arcsec\ zoom in on the target from the final fully mosaiced image which was approximately $17\arcsec\times17\arcsec$ in size.}
    \label{fig:ao_example}
\end{figure}

Palomar observations used a 5-point quincunx dither pattern with steps of 5\arcsec; the dither pattern was performed 3 times -- each dither pattern offset from the previous dither pattern by 0.5\arcsec -- for a total of 15 frames per target. Each night, on-sky flats and darks were obtained. Final flats were obtained from the median average of the dark-subtracted flats.  Sky frames were generated from the median average of the dithered science frames. Each science image was then sky-subtracted and flat-fielded. The reduced science frames were combined into a single mosaiced image, with typical AO-corrected resolutions of 0.1\arcsec.  The sensitivity of the final combined AO images were determined by injecting simulated sources azimuthally around the primary target every $20^\circ $ at separations of integer multiples of the central source's full width at half maximum (FWHM, \citealp{Furlan2017}). The brightness of each injected source was scaled until standard aperture photometry detected it with $5\sigma $ significance.  The final $5\sigma $ limit at each separation was determined from the average of all of the determined limits at that separation and the uncertainty on the limit was set by the rms dispersion of the azimuthal slices at a given radial distance.  A typical reduced image and sensitivity curve for a single target is shown in Figure~\ref{fig:ao_example}.

\subsection{Characterizing Imaged Companions}\label{sec:char}
Out of the 198 imaged systems, we identified 43 with stellar companions through visual inspection; see Section~\ref{sec:sim} for additional work to constrain the companions potentially missed by this search. Each system had measurements in a $K$-band filter ($K$, $K_s$, $Br_\gamma$), and an additional measurement in either $J$ or $H_{\rm cont}$, though we primarily relied on the $K$-band measurements for the remainder of this study. Four stars had multiple nights of observations, in which case we kept data from whichever night's point spread function (PSF) had the smaller FWHM.

To initially characterize our stars, we chose to rely on aperture photometry -- as opposed to PSF photometry -- measurements. While PSF photometry would in principle be more precise and have better sensitivity to close-in companions, the PSF for AO imaging often varies between targets, and the small field-of-view often means there are no nearby reference stars. Furthermore, the vast majority of Kepler/TESS transit discovery papers relied on aperture photometry, so using it again here ensured that we could have the most direct and effective comparison to quantify what had been missed in previous works.

Thus, we used aperture photometry to measure the instrumental magnitudes of each star in each system with a detected companion, where the radii of the star- and sky-annuli were set based on the FWHM of the primary and kept constant for each star in the image. We then measured the on-sky separation (in arcseconds) between each companion and the primary (brightest) star. We discarded objects with separations larger than $\sim$\SI{8}{\arcsecond}, or roughly twice the size of a \kepler pixel, to focus on objects that could not have been resolved through \kepler observations. This separation cut removed the only observed companion(s) in three systems, leaving us with a sample of 40 systems (a companion rate of 20\%) to analyze further.

For the remaining 40 systems, we used the aperture photometry flux measurements to calculate instrumental magnitude differences $(\Delta m_{Ks})$ and analytically propagated photometric uncertainties for each star with respect to the primary. Combining these with 2MASS \citep{Skrutskie2006} magnitudes obtained via the NASA Exoplanet Archive\footnote{\url{https://exoplanetarchive.ipac.caltech.edu/}}, we calculated the de-blended apparent $K$-band magnitudes ($m_{Ks}$) for each star in a given system.

In order to estimate the flux dilution relevant to the \kepler observations, we converted our $m_{Ks}$ magnitudes into $m_{Kp}$ magnitudes in the \kepler band using a Dartmouth isochrone of solar metallicity and age ([Fe/H]=0, age of 4.5\,Gyr; \citealp{Dotter2008}). We removed any points on the isochrone that did not pass the \citet{Huber2016} dwarf star selection criterion:
\begin{equation}\label{eq:dwarf}
    \log{g} > \frac{1}{4.671}\arctan{\left(\frac{T_\mathrm{eff}-6300}{-67.172}\right)}+3.876.
\end{equation}
From the remaining isochrone points, we created a color-magnitude relation by fitting a degree-4 polynomial to the $m_{Kp} - m_{Ks}$ colors and $m_{Ks}$ apparent magnitudes. To convert the absolute magnitudes from the isochrone to apparent magnitudes, we relied on distances derived from \gaia DR3 parallax measurements\footnote{Three stars (KICs 6369366, 9202350, 10678359) lack parallaxes in \gaia DR3, so we adopt their distances as listed in \kepler \texttt{DR25} \citet{Thompson2018}.} \citep{Gaia,GaiaDR3}. We used the resulting color-magnitude polynomials to compute $m_{Kp}$ magnitudes for each star, and verified that our  $m_{Kp}$ magnitudes were consistent with available \gaia magnitudes given the similarity between \kepler and \gaia bandpasses. We then measured $\Delta m_{Kp}$ by subtracting the primary's magnitude from that of each companion.

\begin{figure*}[!htb]
    \centering
    \includegraphics[width=\linewidth]{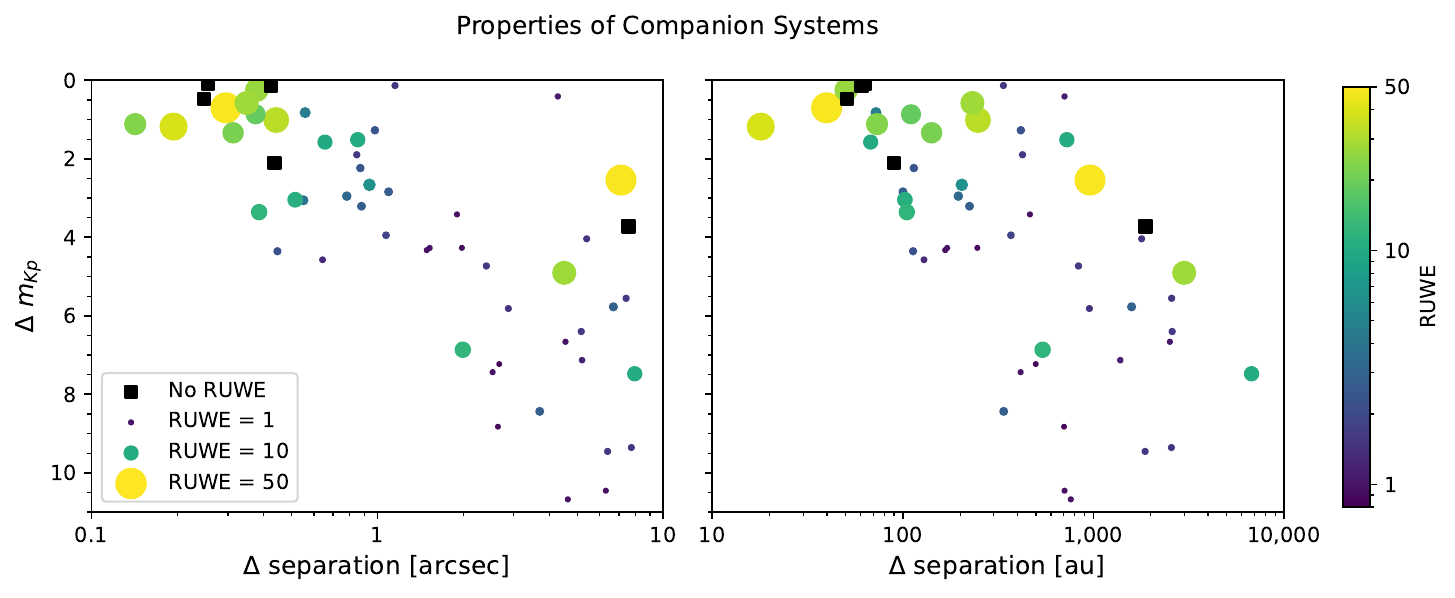}
    \caption{Imaged companions in our \kepler control sample, showing magnitude differences in the \kepler band as functions of projected \textbf{left:} on-sky or \textbf{right:} physical separations (measured with respect to the primary). Circle markers are colored and sized by the RUWE values of the primary star; black square markers indicate systems that did not have RUWE values available in \gaia DR3. The cut-off for on-sky separations to be included in our sample was  roughly twice the \kepler pixel size ($\sim$\SI{8}{\arcsecond}). }
    \label{fig:sample}
\end{figure*}

We also used the distances in conjunction with the on-sky separations to calculate the projected physical separations between companions and primaries; the true separations are likely larger due to misalignment between the system orbital plane and the plane of the sky. The distribution of companion properties is shown in Figure~\ref{fig:sample}, and a table of key companion properties is included Appendix~\ref{app:prop}.

\subsection{Constraining Undetected Companions}\label{sec:sim}
A robust analysis of unresolved stellar companions must account for the companions we observed \textit{and} those we were not sensitive to with our imaging campaign. To identify the population of stellar companions that were not detected in our images, we used code that was originally developed for Palomar and Keck adaptive optics observations \citep{LundCiardi2020AAS...23524906L}. The adapted code is described in depth in \citet{Clark2024AJ....167...56C}, but we outline it briefly here.

The code works by first identifying the population of stellar companions that \textit{could} orbit each star, and then evaluating the sensitivity of each observation to these stellar companions using the derived contrast curves from the high-resolution images. The code creates the populations of stellar companions by first matching the star to a best-fit stellar isochrone from the Dartmouth isochrones \citep{Dotter2008} to estimate the stellar mass range that may have been missed by the high-resolution imaging. The code then draws from the mass ratio and orbital period distributions for solar-type stars from \citet{Raghavan2010ApJS..190....1R} to create the population of simulated stellar companions for each star. We note that while these distributions likely scale with stellar mass, for the FGK mass regime considered in this study, the differences are small and well-represented by an average \citep[see e.g., Figure 2 of][]{Offner2023}. The separations for the simulated companions are calculated assuming a circular orbit, and using the period and stellar masses to compute a semimajor axis. The companion is then given a random position on that orbit. This simulation is performed 10,000 times for each star. These simulations allow us to determine the fraction of stellar companions that would be detectable within each high-resolution image.

To incorporate the effects of the companions that we were \textit{not} sensitive to in our following analysis, we kept simulated companions that (1) would not have been detectable within our imaging sensitivities and (2) had separations less than $\sim$\SI{8}{\arcsecond} (again, roughly twice the \kepler{} pixel size). Many of the remaining ``simulated undetected" companions are within $\sim40$\,au as an approximate boundary for our observation sensitivity (see e.g., the right panel of Figure~\ref{fig:sample}), which is also the approximate peak of the companion semimajor axis distribution for Sun-like stars \citep{Offner2023}.

\subsection{Planet Radius Correction Factors}\label{sec:rcorr}

For each system with a companion (both observed and simulated), we calculated a correction factor, which accounts for the flux contribution from an unresolved stellar companion that would have diluted the transit depth -- and thus the inferred planet radius -- of the original measurement \citep{Ciardi2015}.

If we assume that the planet orbits the primary star, then the correction factor is given by Equation~5 of \citet{Furlan2017}, based on Equation~6 of \citet{Ciardi2015}:
\begin{equation}\label{eq:X1}
    \mathrm{X}_1 = \sqrt{1 + \sum^N_{i=1}{10^{-0.4\Delta m_i}}}.
\end{equation}
where there are $N$ companions in the system, each with a magnitude difference of $\Delta m_i$ with respect to the primary. If one assumes the planet orbits the second-brightest star in the system -- hereafter the ``secondary" -- then the correction factor is instead given by Equation~7 of \citet{Furlan2017}:
\begin{equation}\label{eq:X2}
    \mathrm{X}_2 = \frac{\mathrm{R}_{_\star,2}}{\mathrm{R}_{_\star,1}} \sqrt{10^{0.4\Delta m_c} \left(1 + \sum^N_{i=1}{1 + 10^{-0.4\Delta m_i}}\right)},
\end{equation}
where $\Delta m_c$ is defined as the magnitude of the secondary minus that of the primary. Note that the X$_2$ calculation also requires measurements of stellar radii for the primary and secondary (R$_{_\star,1}$ and R$_{_\star,2}$ respectively). We estimate stellar radii using a degree-4 $m_{Ks}$-R$_\star$ polynomial fit to representative values from the \citet{PecautMamajek} table, filtered to only include F-type stars and smaller (notionally $M_\star < 1.63 M_\odot$). The inferred radii for stars in our control sample are included alongside other key properties in Appendix~\ref{app:prop}.

\begin{figure*}
    \centering
    \includegraphics[width=\linewidth]{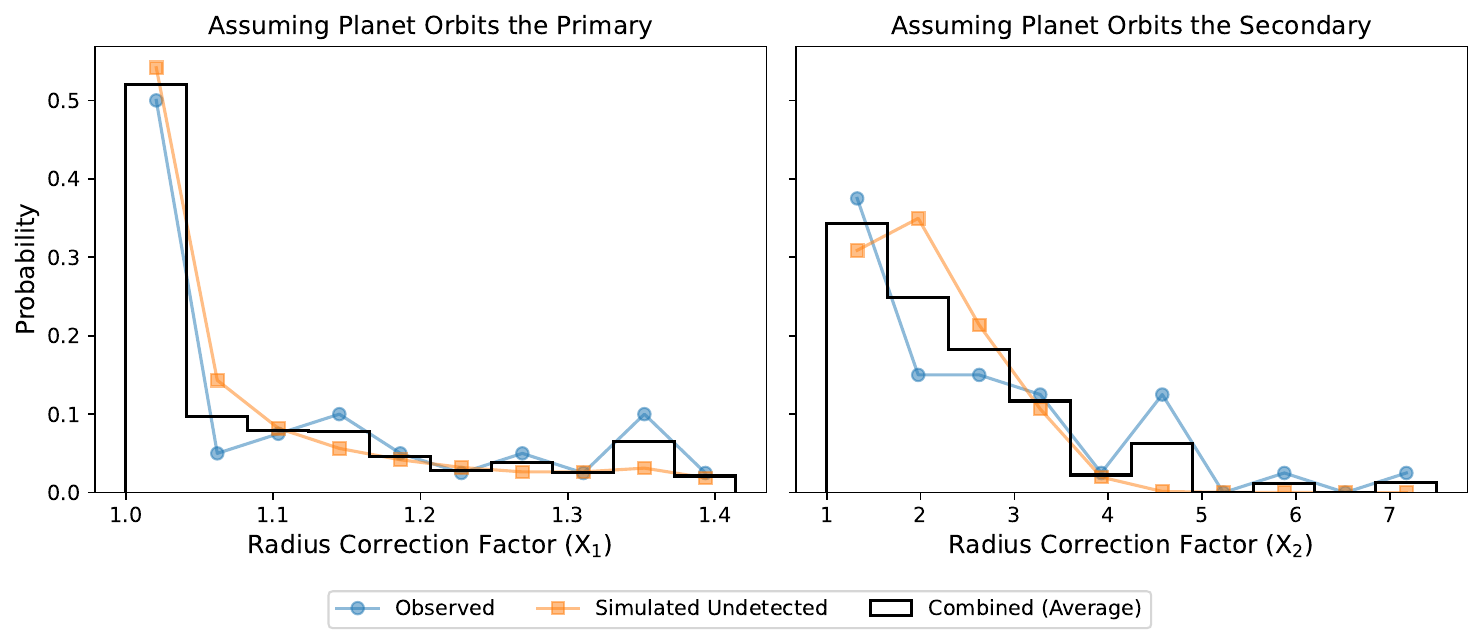}
    \caption{The probability distribution of planet radius correction factors from our \kepler control sample, assuming planets orbit \textbf{left:} the primary or \textbf{right:} the second-brightest star. Blue circles represent the distribution measured from the observed companions, orange squares represent the distribution measured from the simulated, undetected companions, while the black histogram represents a blend of those two distributions averaged in each bin.}
    \label{fig:corrections}
\end{figure*}

The distributions of the resulting correction factors for the systems with observed companions are presented in Figure~\ref{fig:corrections}, along with those for the simulated undetected companions from Section~\ref{sec:sim}. We defined histograms with 10 bins spanning different ranges depending on the assumption of a primary or secondary host star. For primary host stars (X$_1$), bins spanned from 1 to $\sqrt{2}$ (i.e., Equation~\ref{eq:X1} evaluated for two equal-brightness stars); for secondary host stars (X$_2$), bins spanned from 1 to $7.5$ (slightly above our largest value of X$_2$ for an observed system). For each distribution, we then converted the raw counts into a probability density function, and normalized these values by the bin width such that the sum of individual probabilities over each bin would equal 1. To merge the distributions for the observed and simulated companions, we took the average of their probabilities in each bin (separately for X$_1$ and X$_2$), and plotted the corresponding average distribution in Figure~\ref{fig:corrections}. Tables of the raw counts and probabilities are included in Appendix~\ref{app:corr}.


\subsection{Bounding the Unresolved Companion Rate}\label{sec:rates}

Given the prevalence of the simulated undetected companions, incorporating them alongside our observed sample was essential to properly constraining the effect of unresolved companions. But the inclusion of any companions we may have missed required an increase in the companion rate, since there would be more systems with unresolved companions than the 20\% observed in our control sample. While we do not know the true frequency of unresolved companions among \kepler{} stars, one option would be to assume it follows the bias-corrected multiplicity fraction for FGK field stars -- roughly 45\% \citep{Offner2023}. However, this is likely higher than the actual unresolved companion rate for \kepler targets, as many spatially-resolved binaries were already excluded from the Kepler Input Catalog \citep{Brown2011}.

So while the true multiplicity rate remains unknown, from here on we considered two different scenarios to act as lower and upper limits in order to bound how unresolved companions could affect \kepler occurrence rates. In the ``observed" treatment, we used the correction factor distributions and 20\% companion rate observed in our control sample; this is a lower limit knowing that there are companions missed by our imaging campaign. In the ``field" treatment, we used the average of the observed and simulated undetected correction factor distributions at the field star companion rate of 45\%; this is an upper limit knowing that a 45\% multiplicity rate is likely too high, since it includes farther-out companions that had already been spatially resolved and excluded from \kepler stellar catalogs.

\section{Occurrence Rate Modeling with Simulated Companions}\label{sec:occ}

In \citet{Bergsten2022}, we developed a model that described the occurrence distribution of small, close-in planets ($1 - 3.5$\,R$_\oplus$, $2-100$\,days) using two components; please refer to Figure 1 of \citet{Bergsten2022} for a visualization of these functions. First, a broken power law in orbital period described the slopes ($\beta_1$ and $\beta_2$) of the occurrence distributions before and after the break $P_\mathrm{break}$. Second, to constrain the observed relative decrease in super-Earths (and increase in sub-Neptunes) towards larger periods, we modeled the percentage of small planets with super-Earth sizes as a function of orbital period. This component used a hyperbolic tangent to smoothly transition between short- and long-period super-Earth percentages ($\chi_1$ and $\chi_2$ respectively) on either side of a period $P_\mathrm{central}$, with the sharpness of the transition defined by a smoothness parameter $s$. The resulting fractional occurrence curve resembled a predicted effect of atmospheric mass loss, stripping volatile-rich sub-Neptunes down to super-Earth-sized remnant cores at shorter (but not longer) orbital periods \citep[see e.g.,][]{Lopez2018, Pascucci2019}.

Combining the two components together with an overall occurrence parameter $F_0$, the \citet{Bergsten2022} model constrained the occurrence (and fractional occurrence) of super-Earths and sub-Neptunes using eight free parameters $\left(F_0, P_\mathrm{break}, \beta_1, \beta_2, P_\mathrm{central}, s, \chi_1, \chi_2 \right)$. We then extrapolated this model to estimate the occurrence of rocky Earth-sized planets in the habitable zone (notionally $0.7-1.5$\,R$_\oplus$ and $363-811$\,days for Sun-like stars). Notably, \citet{Bergsten2022} operated under the assumption that all stars assumed to be single in the \kepler{} Input Catalog \citep{Brown2011} were indeed single, and did not implement any corrections for unresolved stellar companions.

In this work, we modeled the occurrence rate of small close-in planets \textit{while simulating the presence of stellar companions}. We adopted the same functional form for the occurrence distribution as in \citet{Bergsten2022}, which included a broken power law in orbital period and distributed the occurrence between super-Earths and sub-Neptunes with a period-dependent fraction (with no further occurrence-radius dependencies in either of the two planet classes). We also employed the same \citet{Huber2016} dwarf star criterion of Equation~\ref{eq:dwarf} and adopted the same stellar mass range of $0.56 - 1.63$\,M$_\odot$ to isolate Sun-like (or FGK) stars \citep{PecautMamajek}. We applied these star and planet criteria to the \kepler{} \texttt{DR25} catalog \citep{Thompson2018}, supplemented with \gaia{}-revised stellar properties \citep{Berger2020} and updated planet properties. This left us with a sample of 114,515 FGK dwarf stars hosting 2,062 confirmed and candidate planets within $1 - 3.5$\,R$_\oplus$ and $2-100$\,days (the same parameter range as \citealp{Bergsten2022}). Other than excluding known false positives, we did not impose any other quality cuts to the \texttt{DR25} planet catalog, and accounted for per-candidate reliability in Section~\ref{sec:occ_sim}.

We consider four scenarios, each with their own separate occurrence models. Our first scenario is an update to \citet{Bergsten2022} with key changes described in Appendix~\ref{app:2022updates}, where we establish an occurrence baseline assuming:
\begin{itemize}
    \item Scenario 1: all stars in the sample are single, and all planets in the sample orbit single stars.
\end{itemize}
Then, we expand the \citet{Bergsten2022} framework, using the same functional form to describe the occurrence distribution, but now incorporating and correcting for the presence of unresolved stellar companions. We impose that some presumed-single stars may be unknown multi-star systems, and determine which radius correction factors to invoke depending on whether:
\begin{itemize}
    \item Scenario 2: any planets in multi-star systems orbit the primary (brightest) star. 
    \item Scenario 3: any planets in multi-star systems orbit the secondary (second-brightest) star.
    \item Scenario 4: any planets in multi-star systems may orbit either the primary or secondary in equal proportion.
\end{itemize}
For Scenarios 2--4 involving stellar companions, we fit two separate sets of models (using the methodology described below): one with the rate (20\%) and properties of our observed companions, and one with the field rate (45\%) with a mix of properties from our observed and simulated undetected companions.

\subsection{Model Optimization}\label{sec:occ_sim}

For each scenario, we followed the same methodology of \citet[their Section 2.3]{Bergsten2022} to fit an occurrence rate model to \kepler{} data. In short, we optimized a population distribution function \citep{Youdin2011, Burke2015} by performing $100$ inferences of a Markov Chain Monte Carlo process using \texttt{emcee} \citep{ForemanMackey2012}. In each inference, a model was constrained to a population of \kepler{} planets which were drawn from our initial sample in proportion to their reliability -- e.g., a $10\%$ reliable candidate is included in $10\%$ of inferences; see \citet{Bryson2020-OG-Reliability, Bryson2020, Bryson2021}. Once each inference of 64 walkers ran for 10,000 steps, we discarded the first 1,000 steps from each inference for burn-in.\footnote{Optimization criteria were based on measurements of typical auto-correlation time $\tau$, running models for at least $50\tau$ and discarding at least $2\tau$ for burn-in.} Finally, we concatenated every inference's posteriors to represent the reliability-weighted posterior distribution of model free parameters. 

For the three scenarios involving stellar companions, we also included a simulation of how unresolved stellar companions could affect completeness measurements. Again, this analysis was performed in two separate treatments, using either the observed companion rate (20\%) and properties, or the field rate (45\%) with a mixture of observed and simulated undetected companion properties. In each modeling inference, each \kepler{} star had a chance (either 20 or 45\% depending on the treatment) of being assigned a stellar companion. For systems assigned a companion, a radius correction factor was then drawn from the distributions in Figure~\ref{fig:corrections} (using the observed or combined distributions for the observed- and field-rate treatments, respectively) depending on the relevant scenario:
\begin{itemize}
    \item Scenario 2: planet orbits the primary, with all correction factors drawn from the $X_1$ distribution.
    \item Scenario 3: planet orbits the secondary, with all correction factors drawn from the $X_2$ distribution.
    \item Scenario 4: planet has an equal chance of orbiting the primary or secondary, so a given system's correction factor has a $50\%$/$50\%$ chance of being drawn from the $X_1$/$X_2$ distributions, respectively.
\end{itemize}

We note here that Scenario 4 assumes planets are equally likely to orbit the primary or secondary. The true host distribution may be skewed towards primaries if planets form more often around primaries, which tend to have more massive protoplanetary disks \citep[see e.g.,][]{AkesonJensen2014}. Conversely, secondaries may be the more likely hosts if the trend of short-period small planets being more common around lower-mass stars \citep{DressingCharbonneau2015, Bergsten2023} also holds at longer (i.e., typical companion separation) orbital periods. An alternative path would be the weighted average approach of \citet{Furlan2017}, in which the frequency of correction factors was instead drawn from a blending of their $X_1$ and $X_2$ distributions at some unequal proportion that favored the primary. Because observational constraints for the frequency at which planets orbit the primary/secondary remain unclear, we leave this opportunity for future works. 

\begin{figure*}[!htb]
    \centering
    \minipage{\linewidth}
      \includegraphics[width=\textwidth]{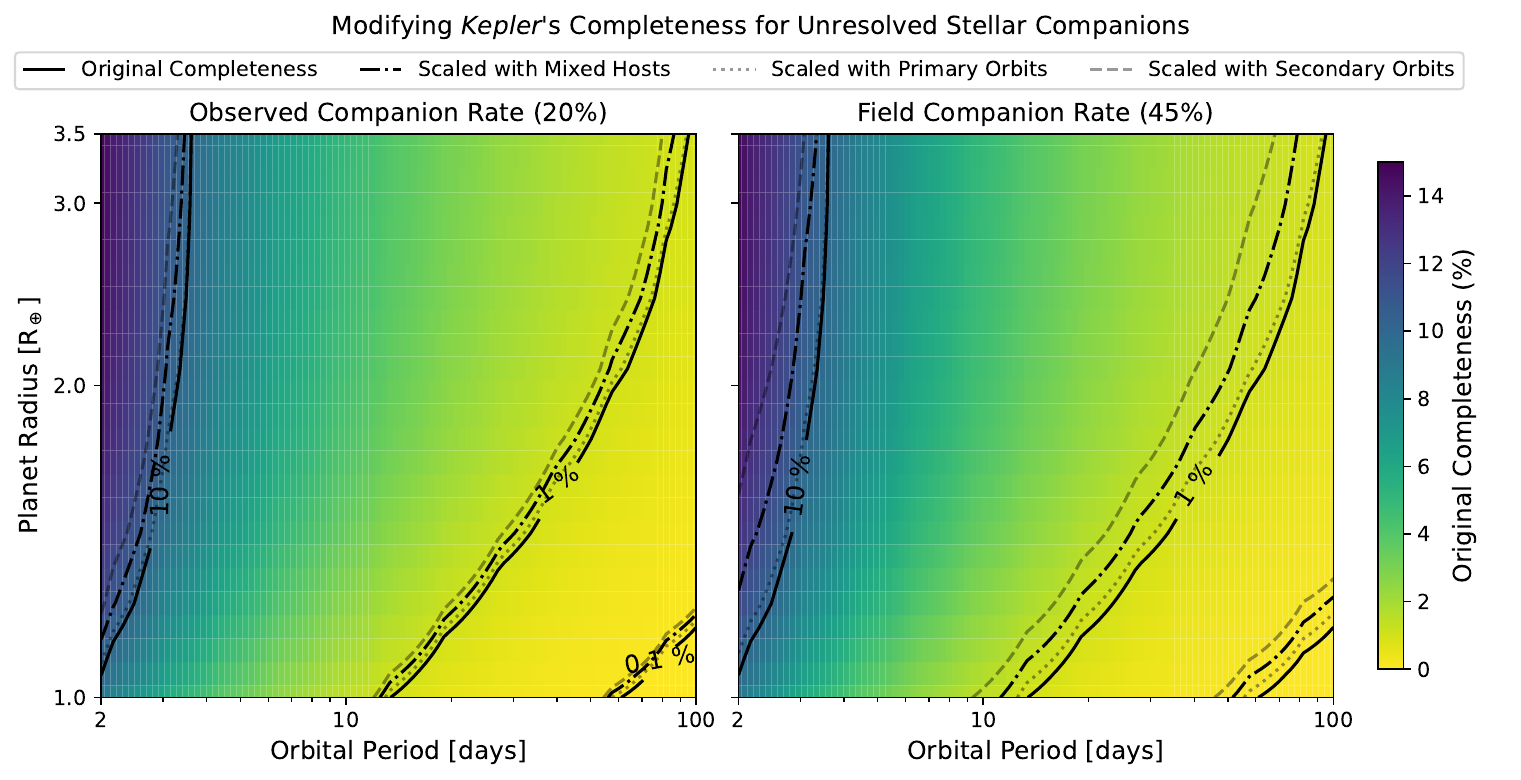}
    \endminipage\hfill
    \minipage{\linewidth}
      \includegraphics[width=\textwidth]{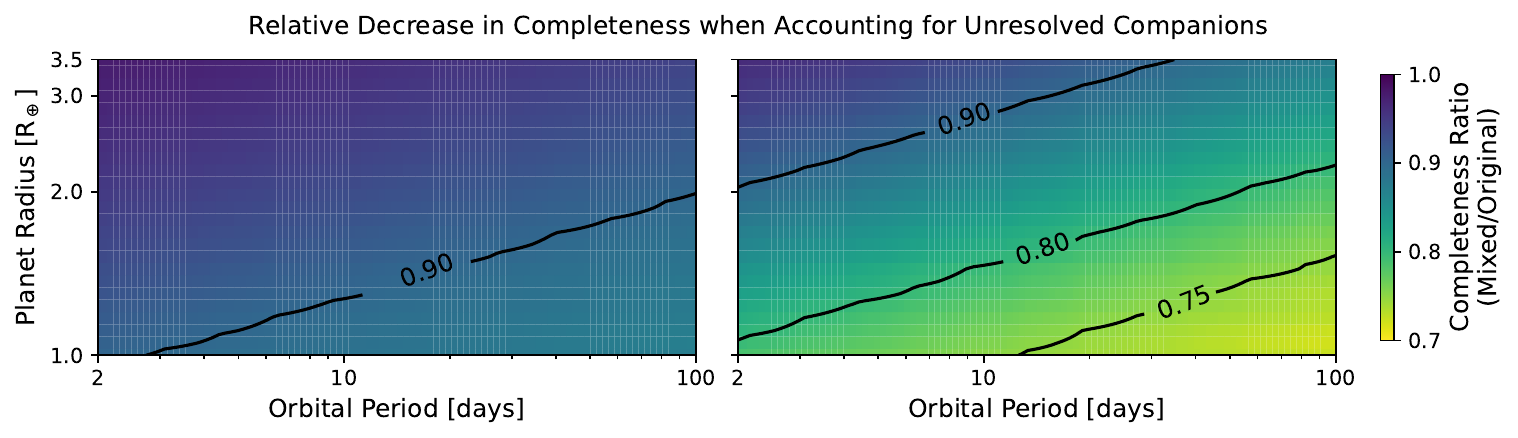}
    \endminipage\hfill
    \caption{Modifications to the \kepler{} survey completeness for our sample, based on the simulated presence of stellar companions using \textbf{left:} the properties of our observed companion sample at their observed rate of 20\% or \textbf{right:} a combination of properties from our observed and simulated undetected companions at the field rate of 45\%. \textbf{Top:} the colormap denotes the original (i.e., Scenario 1) \kepler{} average survey completeness, with black solid contours at the 0.1, 1, and $10\%$ levels. Other contours represent the average completeness maps from scenarios where some stars are assigned companions, shifting their completeness grids vertically upwards. These shifts depend on radius correction factors drawn assuming planets orbit the primary (dotted, Scenario 2), secondary (dashed, Scenario 3), or a mix of the two (dot-dashed, Scenario 4). \textbf{Bottom:} the colormap denotes the ratio of average completeness maps with and without companions (Scenario 4 / Scenario 1). Contours show where the Scenario 4 map equals either 75, 80, or $90\%$ of the original Scenario 1 map.}
    \label{fig:comp}
\end{figure*}

After the radius correction factor was drawn, the completeness grid for the star assigned a companion was modified by multiplying the planet radius axis by the drawn correction factor. This induces a population-level effect where \kepler{}'s sensitivity to detecting planets of a given radius is lowered in systems with unresolved stellar companions, in proportion to how much \kepler{} would have underestimated planet radii in those systems. Critically, we assume the measured radii of \kepler{} planets are ``true" -- and do not adjust them by the same correction factors -- in order to isolate how the specific effect of unresolved companions on detection sensitivity can generally impact planet occurrence rates. While the sensitivity adjustments are handled per-star, the population-averaged effect is visualized in Figure~\ref{fig:comp}. For stars without companions in a given inference (100\% of stars in Scenario 1, and either 80\% or 55\% of stars in Scenarios 2--4 depending on the treatment), the completeness maps were not shifted along the radius axis.

Each star's completeness map was then interpolated onto a common grid used for model evaluation, and these grids were added together, creating the summed completeness grids necessary for evaluating the likelihood function (see e.g., Equation 10 of \citealp{Burke2015}) used when optimizing the population distribution function; Figure~\ref{fig:comp} presents the average version of these grids. The final completeness maps are dependent on (a) which stars are randomly assigned companions, and (b) the correction factors randomly assigned to those companion systems. As such, there are $601$ (potentially) unique completeness maps generated in this work: 1 for each of the 100 inferences in Scenarios 2-4 -- created separately for the observed and field treatments --  and only 1 map used for all 100 inferences of Scenario 1.

\section{Results}\label{sec:results}

With these optimized model posterior distributions, we calculated planet occurrence rates by evaluating the \citet{Bergsten2022} population distribution over a given range of orbital periods and planet radii. Here we explore two specific cases based on our motivations from Section~\ref{sec:intro}: how does the presence of unresolved companions affect small planet occurrence rates, and to what degree does correcting for unresolved companions impact occurrence rates in the habitable zone?

\subsection{Small, Close-in Planets}\label{sec:small}

\begin{figure*}
    \centering
    \includegraphics[width=\textwidth]{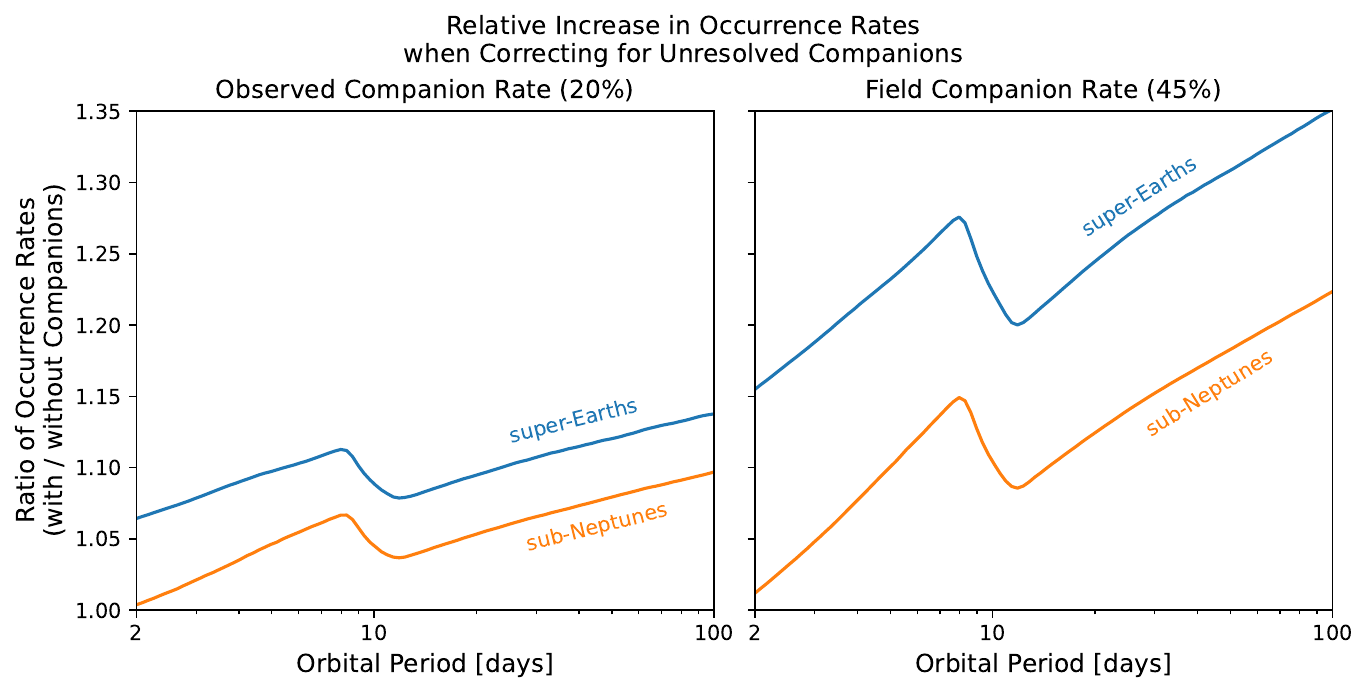}
    \caption{The ratio of median occurrence rates from our models with and without a treatment for unresolved stellar companions (Scenarios 4/1), using \textbf{left:} the observed companion rate and properties, and \textbf{right:} the field companion rate and blend of observed/simulated undetected properties. The largest changes are for far-out super-Earths (blue; 1--2\,R$_\oplus$) and the smallest changes are for close-in sub-Neptunes (orange; 2--3.5\,R$_\oplus$). The kink in these ratios around $\sim$10~days is due to a slight difference in the best-fit orbital period-occurrence distributions (Fig.~\ref{fig:occ_marg}), discussed in Section~\ref{sec:small}.}
    \label{fig:occ_ratio}
\end{figure*}

\begin{figure*}
    \centering
    \includegraphics[width=\textwidth]{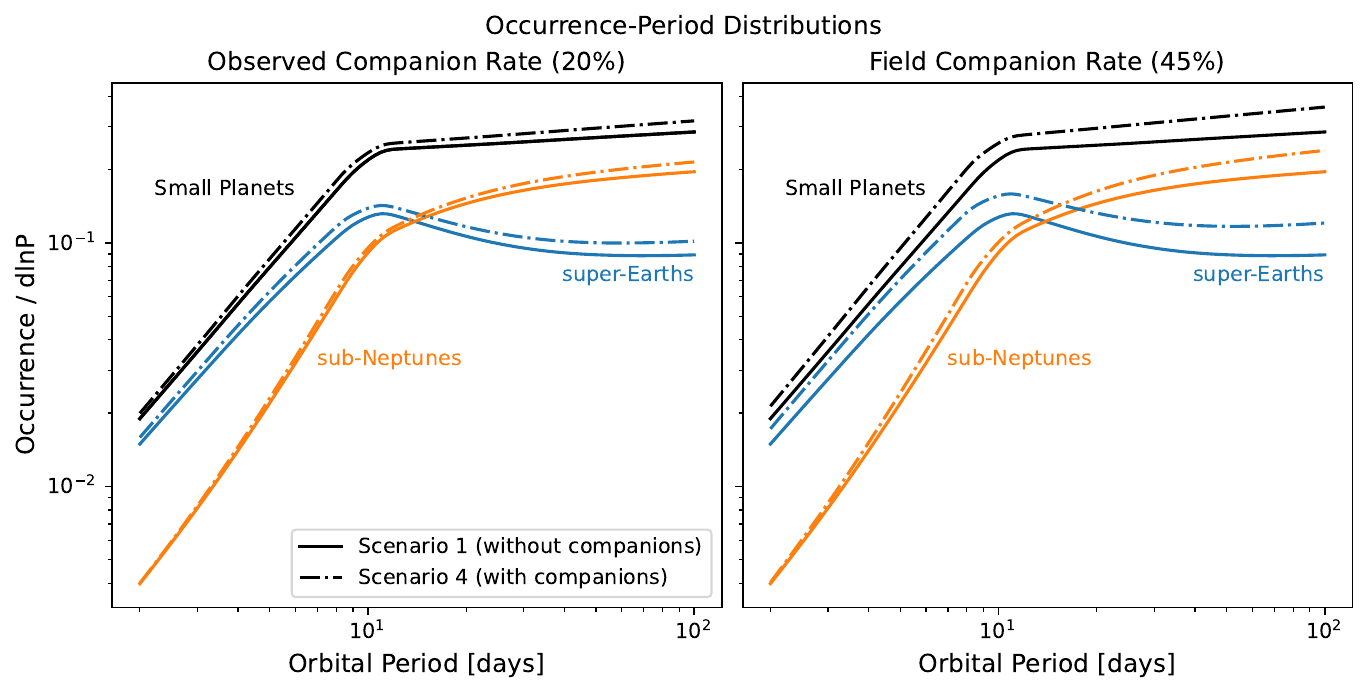}
    \caption{The median occurrence-period distributions from our models with (Scenario 4; dot-dashed lines) and without (Scenario 1; solid lines) a treatment for unresolved stellar companions, using either \textbf{left:} the observed companion rate and properties, or \textbf{right:} the field companion rate and blend of observed/simulated undetected properties. Colors denote the marginalized distributions for either super-Earths (blue; 1--2\,R$_\oplus$), sub-Neptunes (orange; 2--3.5\,R$_\oplus$) or their combination for all small planets (black; 1--3.5\,R$_\oplus$).}
    \label{fig:occ_marg}
\end{figure*}

We evaluated our model's median occurrence rates over a range of orbital periods and planet radii for three cases: Scenario 1, with no corrections for unresolved stellar companions; Scenario 4 in the observed treatment (20\% companion rate using the observed companion properties, with equal odds of planets orbiting a primary or secondary star); and Scenario 4 in the field treatment (45\% companion rate using the combined observed and simulated undetected companion properties).  We chose to focus on Scenario 4 as a representative case for the scenarios that correct for unresolved stellar companions. We then calculated the ratio of median occurrence rates from Scenarios 4/1 (with / without corrections for unresolved companions) in both the observed and field treatments; the resulting ratios as a function of orbital period are shown in Figure~\ref{fig:occ_ratio}. An example of the underlying occurrence-period distributions is shown in Figure~\ref{fig:occ_marg}.

Our methodology of scaling the completeness contours by a radius correction factor necessarily means that occurrence rates will, on average, be higher when accounting for stellar companions, since radius correction factors can never be less than unity. However, we note that if the change in occurrence from incorporating this companion treatment is small compared to the intrinsic uncertainty in the occurrence model, it is possible for a random draw from the ``corrected" models to be a lower occurrence than the ``uncorrected" models. Additionally, \kepler's completeness contours are curved in the period-radius plane (e.g., typically more shallow at smaller planet radii and at longer orbital periods; see Figure~\ref{fig:comp}). In turn, the vertical shift of these contours along the radius axis in Scenarios 2-4 means that the difference in completeness from the unshifted contours of Scenario 1 are larger for (a) smaller planets and (b) longer-period planets (see e.g., the lower panels of Figure~\ref{fig:comp}).

Consequently, this same trend also applies to the ratio of occurrence rates between Scenarios 2-4 and Scenario 1,  as demonstrated in Figure~\ref{fig:occ_ratio}. The \citet{Bergsten2022} model's only dependence on planet radius is the distinction between super-Earths (1--2\,R$_\oplus$) and sub-Neptunes (2--3.5\,R$_\oplus$)\footnote{The split at $\sim$2\,R$_\oplus$ from \citet{Bergsten2022} is based on the stellar-mass dependent radius valley scaling from \citealp{Wu2019}.} with a flat (in log space) occurrence-radius distribution within each bin. As such, the completeness-radius effect described previously is generalized into those two bins in the resulting occurrence rates (despite models being evaluated at a higher radius resolution).

\begin{table*}
\centering
    \begin{tabular}{|l|c|cc|c|}
    \hline
         Companion Treatment&  Planet Radius Range&  Minimum Ratio&  Maximum Ratio& Ratio Averaged over log-Period\\
 & [R$_\oplus$]& (at 2\,days)& (at 100\,days)&\\ \hline\hline
         \multirow{2}{*}{Observed Rate (20\%)}&  $1-2$&  $1.06_{-0.16}^{+0.19}$ &  $1.14_{-0.26}^{+0.33}$ & $1.10_{-0.13}^{+0.15}$\\
         &  $2-3.5$&  $1.00_{-0.21}^{+0.26}$&  $1.10_{-0.15}^{+0.17}$ & $1.06_{-0.11}^{+0.12}$\\
         \hline
         \multirow{2}{*}{Field Rate (45\%)}&  $1-2$&  $1.16_{-0.18}^{+0.21}$&  $1.35_{-0.31}^{+0.40}$& $1.24_{-0.16}^{+0.19}$\\
         &  $2-3.5$&  $1.01_{-0.21}^{+0.27}$&  $1.22_{-0.16}^{+0.19}$& $1.13_{-0.13}^{+0.13}$\\
    \hline
    \end{tabular}
    \caption{Ratio of occurrence rates with/without corrections for unresolved companions; please refer to Figure~\ref{fig:occ_ratio} for the distribution of these ratios as a function of orbital period. Quoted values use the ratio of median and \nth{16}/\nth{84} percentile occurrence rates from Scenario 4 (mixture of primary and secondary hosts) relative to Scenario 1 (no corrections).}
    \label{tab:occ_ratio}
\end{table*}

The occurrence ratio (with/without accounting for stellar companions) for super-Earths is always larger than that for sub-Neptunes. In the observed treatment, the ratio for super-Earths grows slightly from $1.06_{-0.16}^{+0.19}$ at 2~days to $1.14_{-0.26}^{+0.33}$ at 100~days (averaging over log-period to $1.10_{-0.13}^{+0.15}$), while the ratio for sub-Neptunes grows from $1.00_{-0.21}^{+0.26}$ to $1.10_{-0.15}^{+0.17}$ over the same range (averaging to $1.06_{-0.11}^{+0.12}$). Similar trends are amplified in the field treatment, with the ratio for super-Earths rising from $1.16_{-0.18}^{+0.21}$ to $1.35_{-0.31}^{+0.40}$ (avg.\,$1.24_{-0.16}^{+0.19}$) and sub-Neptunes rising from $1.01_{-0.21}^{+0.27}$ to $1.22_{-0.16}^{+0.19}$ (avg.\,$1.13_{-0.13}^{+0.13}$). Averaged over the entire domain (1--3.5\,R$_\oplus$, 2--100\,days), the median occurrence rate of small, close-in planets increases when correcting for stellar companions by a factor of $1.08_{-0.12}^{+0.14}$ in the observed treatment and $1.18_{-0.14}^{+0.17}$ in the field treatment. These values are summarized in Table~\ref{tab:occ_ratio}.

Figure~\ref{fig:occ_ratio} also shows a kink in the median occurrence ratios around $\sim$10~days where the ratio briefly decreases, before continuing to increase towards larger orbital periods. This corresponds to a difference in free parameters between the models of Scenarios 1 and 4, where the latter exhibits a slightly earlier break in the occurrence-period power law compared to the former (e.g., ${10.28}^{+1.24}_{-1.75}$ vs ${9.64}^{+1.53}_{-1.79}$\,days in the field treatment). This offset introduces a region where the occurrence distribution in Scenario 1 continues to increase while the one in Scenario 4 begins to flatten out (see e.g., the right panel of Fig.~\ref{fig:occ_marg}), meaning the ratio between values briefly decreases. Once the distribution in Scenario 1 also flattens out, the ratio returns to a trend of increasing towards larger periods due to models with companions having steeper occurrence-period slopes than those without. We note that the difference between these break parameters is not statistically significant, and not something indicative of any underlying physical processes shaping the population differently for planets in single-/multi-star systems. A table of the median and \nth{16}/\nth{84}-percentile best-fit model parameters is included in Appendix~\ref{app:params}.

\subsection{Habitable-Zone Occurrence Rates}

\begin{table*}
    \centering
    \begin{tabular}{|c|c|c|c|c|}
    \hline
  & Description& \multicolumn{2}{c|}{$\eta_\oplus$ [$\%$]} & Increase (Relative\\  &  & Conservative&Optimistic& to No Companions) \\ \hline \hline
          &  No treatment of stellar companions& ${8.4}_{-3.2}^{+2.2}$ & ${13.3}_{-4.8}^{+3.4}$& ${--}$ \\ \hline
          \multirow{3}{*}{Observed Rate (20\%)}&  Planets orbit primary& ${9.2}_{-3.6}^{+2.5}$ & ${14.6}_{-5.5}^{+3.9}$&${1.10}_{-0.62}^{+0.40}$ \\ 
          &  Planets orbit secondary& ${10.8}_{-4.3}^{+3.0}$ & ${17.1}_{-6.4}^{+4.6}$&${1.29}_{-0.72}^{+0.47}$ \\ 
          &  Mix of primary/secondary orbits& ${9.9}_{-3.9}^{+2.8}$ & ${15.7}_{-5.9}^{+4.2}$&${1.18}_{-0.66}^{+0.43}$ \\\hline
          \multirow{3}{*}{Field Rate (45\%)}& Planets orbit primary& ${9.9}_{-3.9}^{+2.8}$ & ${15.7}_{-5.9}^{+4.2}$ &${1.18}_{-0.66}^{+0.43}$\\
          & Planets orbit secondary& ${16.0}_{-6.1}^{+4.6}$ & ${25.1}_{-9.1}^{+7.0}$ & ${1.89}_{-1.05}^{+0.69}$\\
          & Mix of primary/secondary orbits& ${12.3}_{-4.8}^{+3.5}$ & ${19.4}_{-7.2}^{+5.3}$ &${1.46}_{-0.83}^{+0.53}$\\
    \hline
    \end{tabular}
    \caption{Updated estimates for $\eta_\oplus$. Values are derived by extrapolating and integrating our model for Earth-sized (0.7--1.5\,R$_\oplus$) planets in the habitable zone (based on \citealp{Kopparapu2013}).}
    \label{tab:eta}
\end{table*}

Though our models are constrained on planets within 100 days, we can follow the standard procedure of extrapolating them to longer periods in order to estimate the occurrence of Earth-sized planets in the habitable zone ($\eta_\oplus$). We used 0.7--1.5\,R$_\oplus$ as our definition of ``Earth-sized" planets. For the habitable zone, we adopted two definitions in orbital period based on the limits from \citet{Kopparapu2013} evaluated for the average temperature ($\sim$5665\,K) of our \kepler FGK sample: 363--811\,days for a conservative habitable zone (between the moist- and maximum-greenhouse limits), and 238--867\,days for an optimistic habitable zone (between the recent Venus and early Mars limits).

For each scenario, we produced a distribution of $\eta_\oplus$ estimates by taking each parameter vector from the MCMC posteriors and integrating the \citet{Bergsten2022} model over the aforementioned period and radius ranges. The median and lower/upper $1\sigma$ estimates (\nth{50} and \nth{16}/\nth{84} percentiles respectively) of the $\eta_\oplus$ distributions from each scenario are included in Table~\ref{tab:eta}. We also took the ratio of the $\eta_\oplus$ distributions from Scenarios 2-4 divided by the distribution from Scenario 1 to represent the relative increase in $\eta_\oplus$ that arises when including a treatment of unresolved stellar companions; these values are included in Table~\ref{tab:eta}.\footnote{We quote the $\eta_\oplus$ ratios from the conservative habitable zone case, which had slightly larger uncertainties than those from the optimistic case.} We reiterate that, while our methodology is such that the occurrence rates from Scenarios 2-4 would exceed Scenario 1 for a given model parameter vector, the uncertainty on the parameter vectors is wide enough that a random $\eta_\oplus$ instance from the latter could exceed the former. As such, we note that the relative ``increases" in Table~\ref{tab:eta} can be $< 1$ (i.e., a decrease) at $1\sigma$, but emphasize that such decreases stem from the inherent uncertainty in the occurrence model (see e.g., a similar problem in Table~\ref{tab:occ_ratio}) and not the effects of unresolved companions. Despite this, an increased occurrence is favored in the majority (60--90\% depending on the treatment/scenario) of iterations.

For Scenario 1 without stellar companions (an update of \citealp{Bergsten2022}), we found  $\eta_\oplus = {8.4}_{-3.2}^{+2.2}\%$ for a conservative habitable zone\footnote{\citet{Bergsten2022} originally reported $\eta_\oplus = {9.4}_{-2.5}^{+3.4}\%$ for the conservative habitable zone. The difference in this work is due to our use of per-star instead of sample-averaged completeness maps, described in Appendix~\ref{app:2022updates}.} and ${13.3}_{-4.8}^{+3.4}\%$ for an optimistic one (which was not originally considered in \citealp{Bergsten2022}). For our observed treatment with a mix of planet radius correction factors from primary and secondary host stars (Scenario 4), we found $\eta_\oplus = {9.9}_{-3.9}^{+2.8}\%$ and ${15.7}_{-5.9}^{+4.2}\%$ for the same habitable zone definitions; in the field treatment with Scenario 4, we found $\eta_\oplus = {12.3}_{-4.8}^{+3.5}\%$ and ${19.4}_{-7.2}^{+5.3}\%$.

Values from Scenario 4 are ${1.18}_{-0.66}^{+0.43}$ times larger than those from Scenario 1 in the observed treatment, and ${1.46}_{-0.83}^{+0.53}$ times larger in the field treatment. As discussed in Section~\ref{sec:rates}, the true prevalence of unresolved companions is bounded by these two treatments: first, the observed treatment (with a companion rate of 20\%) represents a lower limit due to incomplete sensitivity of our observations. Second, the field treatment (with a companion rate of 45\%) is an upper limit since the removal of spatially-resolved binaries from the KIC should means the remaining rate of unresolved companions should be lower than the FGK field star multiplicity rate. Following this interpretation, correcting for unresolved stellar companions can increase $\eta_\oplus$ measurements by some factor bounded between ${1.18}_{-0.66}^{+0.43}$ and ${1.46}_{-0.83}^{+0.53}$ (or, at their median values, an increase between $\sim$1.2--1.45) times the uncorrected value.\footnote{The similarity between the median increases and the companion rates used (roughly 1.2 for 20\%, and 1.45 for 45\%) is a coincidence based on the integration parameters for our definition of Earth-like planets. Adopting different period or radius integration bounds would yield different increases (see e.g., Figure~\ref{fig:occ_ratio}).} Again, the overlapping limits reflect current uncertainties in the underlying occurrence models -- and the shortcomings of their extrapolation -- which propagate and compound when estimating $\eta_\oplus$.

\section{Conclusions}\label{sec:end}

We conducted a high-resolution imaging survey on a control sample of \kepler{} stars. We then incorporated our measurements of unresolved stellar companions into an occurrence rate model for close-in small planets with separate components for super-Earths and sub-Neptunes. We performed our analysis for two separate treatments: one with the rate (20\%) and properties of our observed sample, and one combining their properties with those of simulated undetected companions at the field companion rate (45\%). These treatments effectively bound the influence of unresolved stellar companions on \kepler{} occurrence rate calculations, quantified with two key results: 

\begin{itemize}
    \item We find that small planet occurrence rates between 2--100~days increase by an average factor of 1.08 or 1.19 depending on the companion treatment, with the largest increases felt by smaller planets at larger orbital periods.
    
    \item We extrapolate our model to estimate the occurrence of Earth-sized planets in the habitable zone ($\eta_\oplus$). We determine that $\eta_\oplus$ can increase by a factor of ${1.18}_{-0.66}^{+0.43}$ -- ${1.46}_{-0.83}^{+0.53}$ (from ${13.3}_{-4.8}^{+3.4}\%$ to ${15.7}_{-5.9}^{+4.2}$ -- ${19.4}_{-7.2}^{+5.3}\%$ for an optimistic habitable zone) when accounting for unresolved stellar companions.
\end{itemize}

The new, larger $\eta_\oplus$ values are closer to the $24^{+46}_{-16}\%$ value assumed for the LUVOIR \citep{LUVOIR} and HabEx \citep{HabEx} mission concept studies. The estimate of $\eta_\oplus \approx 24\%$ \citep[stemming from][]{Kopparapu2018} formed the basis of the \citet{Decadal} decadal survey recommendation for a telescope to detect and characterize 25 nearby Earth-like planets, spurring the development of the Habitable Worlds Observatory (HWO). The original \citet{Kopparapu2018} study did not account for the atmospheric evolution of small close-in planets which drove $\eta_\oplus$ estimates lower \citep{Bergsten2022}, nor did it account for the presence of unresolved stellar companions affecting \kepler{}'s detection sensitivity which drove $\eta_\oplus$ estimates higher. While we have accounted for both of the above here, it is important to note that we did not incorporate information about the multiplicity of planet hosts to modify \kepler{} planet radii, which would likely have a competing effect to drive occurrence rates down. Such considerations would be a necessary step for future works attempting to providing true ``multiplicity-corrected" occurrence rate estimates (going beyond our focus on the detection sensitivity issue).

Furthermore, another challenge remains for HWO and stellar multiplicity: namely, the demographics of planets in multi-star systems. This work is but one addition to the study of how stellar companions affect measurements of planet occurrence rates \citep[e.g.,][]{Hirsch2017AJ, Bouma2018, Teske2018, savel2020, Moe2021, Sullivan2022a, Sullivan2022b}, but comparatively fewer studies have pursued the demographics of planets specifically in known multi-star systems \citep[e.g.,][]{Hirsch2021, Sullivan2024}. Yet because HWO is likely to observe some nearby multi-star systems in its search for Earth-like planets \citep{Mamajek2024, Tuchow2024}, demographic studies (and potentially $\eta_\oplus$ estimates) tailored to multi-star systems remain a critical area of study for future works. Furthermore, such studies enable comparisons between single- and multi-star system planet demographics \citep{Sullivan2023, Clark2024}, which furthers our understanding of how these populations form and evolve. As such, despite the many difficulties stellar companions introduce (to planet detection, characterization, and demographics), they represent an important consideration for future works that should not be neglected.

\section{Acknowledgements}

This work was supported in part through a Visiting Graduate Student Fellowship (VGSF) at Caltech/IPAC. G.J.B. would like to thank Samantha N. Hasler, Mason D. Ruby, Grant P. Donnelly, and Rachel B. Fernandes for lending their support and expertise throughout this work.

G.J.B. and I.P. acknowledge partial support from NASA grant 23-XRP23\_2-0095.
D.R.C. acknowledges partial support from NASA Grant 18-2XRP18\_2-0007.
I.P. acknowledges partial support from NASA under agreement No. 80NSSC21K0593 for the program “Alien Earths.”

This research has made use of the NASA Exoplanet Archive, which is operated by the California Institute of Technology, under contract with the National Aeronautics and Space Administration under the Exoplanet Exploration Program. This work is based on observations obtained at the Hale Telescope, Palomar Observatory, as part of a collaborative agreement between the Caltech Optical Observatories and the Jet Propulsion Laboratory, operated by Caltech for NASA. The results reported herein benefited from collaborations and/or information exchange within NASA’s Nexus for Exoplanet System Science (NExSS) research coordination network sponsored by NASA’s Science Mission Directorate.

This work has made use of data from the European Space Agency (ESA) mission {\it Gaia} (\url{https://www.cosmos.esa.int/gaia}), processed by the {\it Gaia} Data Processing and Analysis Consortium (DPAC, \url{https://www.cosmos.esa.int/web/gaia/dpac/consortium}). Funding for the DPAC has been provided by national institutions, in particular the institutions participating in the {\it Gaia} Multilateral Agreement.

\vspace{5mm}
\facilities{\gaia{}, \kepler{}}

\software{\texttt{NumPy} \citep{numpy}, \texttt{SciPy} \citep{scipy}, \texttt{Matplotlib} \citep{pyplot}, \texttt{emcee} \citep{ForemanMackey2012}, \texttt{corner} \citep{corner}, \texttt{epos} \citep{Mulders2018}, \texttt{KeplerPORTs} \citep{BurkeCatanzarite2017}}

\bibliography{ref}{}
\bibliographystyle{aasjournal}

\appendix

\section{Properties of Companion Systems in Control Sample}\label{app:prop}

Table~\ref{tab:comps} provides information about each system in our control sample for which companions within 8\arcsec were detected. Each entry includes separation and magnitude estimates determined from our imaging observations and \gaia DR3 data \citep{Gaia,GaiaDR3} as described in Section~\ref{sec:char}, alongside estimates for stellar radii using an $m_{Ks}$-R$_\star$ polynomial based on \citet{PecautMamajek} as described in Section~\ref{sec:rcorr}.

\begin{longtable*}{|ccccccc|}
\hline
         KIC & RUWE & Separation (\SI{}{\arcsec}) & $\Delta m_K$ & Deblended $m_K$ & err($m_K$) & R$_\star$ [R$_\odot$]\\ \hline
         ${3096995}$ &  ${2.37}$ & ${--}$ & ${--}$ &  ${8.74}$ &  ${0.02}$ &  ${1.05}$ \\
         & ${--}$ &  ${0.87}$ &  ${1.26}$ &  ${10.01}$ &  ${0.02}$ &  ${0.69}$ \\ \hline
        ${3640967}$ &  ${48.22}$ & ${--}$ & ${--}$ &  ${11.39}$ &  ${0.02}$ &  ${0.47}$ \\
       ${--}$ & ${--}$ &  ${0.30}$ &  ${0.50}$ &  ${11.89}$ &  ${0.02}$ &  ${0.41}$ \\
       ${--}$ & ${--}$ &  ${7.12}$ &  ${2.18}$ &  ${13.57}$ &  ${0.03}$ &  ${0.24}$ \\ \hline
        ${3648376}$ &  ${4.58}$ & ${--}$ & ${--}$ &  ${10.97}$ &  ${0.02}$ &  ${0.52}$ \\
       ${--}$ & ${--}$ &  ${0.56}$ &  ${0.54}$ &  ${11.51}$ &  ${0.02}$ &  ${0.45}$ \\ \hline
        ${4175216}$ &  ${0.94}$ & ${--}$ & ${--}$ &  ${11.44}$ &  ${0.02}$ &  ${0.67}$ \\
       ${--}$ & ${--}$ &  ${1.90}$ &  ${2.24}$ &  ${13.69}$ &  ${0.02}$ &  ${0.36}$ \\ \hline
        ${4555243}$ &  ${23.38}$ & ${--}$ & ${--}$ &  ${12.45}$ &  ${0.02}$ &  ${0.82}$ \\
       ${--}$ & ${--}$ &  ${0.14}$ &  ${0.61}$ &  ${13.06}$ &  ${0.02}$ &  ${0.67}$ \\ \hline
        ${4906883}$ &  ${1.54}$ & ${--}$ & ${--}$ &  ${10.13}$ &  ${0.02}$ &  ${1.20}$ \\
       ${--}$ & ${--}$ &  ${1.15}$ &  ${0.10}$ &  ${10.23}$ &  ${0.02}$ &  ${1.16}$ \\
       ${--}$ & ${--}$ &  ${6.39}$ &  ${6.62}$ &  ${16.75}$ &  ${0.20}$ &  ${0.12}$ \\ \hline
        ${5351968}$ &  ${32.75}$ & ${--}$ & ${--}$ &  ${11.68}$ &  ${0.02}$ &  ${1.14}$ \\
       ${--}$ & ${--}$ &  ${0.44}$ &  ${0.63}$ &  ${12.31}$ &  ${0.02}$ &  ${0.91}$ \\ \hline
        ${5384713}$ &  ${2.71}$ & ${--}$ & ${--}$ &  ${10.43}$ &  ${0.02}$ &  ${0.49}$ \\
       ${--}$ & ${--}$ &  ${1.10}$ &  ${2.41}$ &  ${12.84}$ &  ${0.02}$ &  ${0.23}$ \\
       ${--}$ & ${--}$ &  ${3.70}$ &  ${5.14}$ &  ${15.57}$ &  ${0.03}$ &  ${0.11}$ \\ \hline
        ${5772806}$ &  ${6.38}$ & ${--}$ & ${--}$ &  ${11.00}$ &  ${0.03}$ &  ${0.71}$ \\
       ${--}$ & ${--}$ &  ${0.94}$ &  ${1.58}$ &  ${12.58}$ &  ${0.03}$ &  ${0.45}$ \\ \hline
        ${6184133}$ &  ${1.36}$ & ${--}$ & ${--}$ &  ${10.31}$ &  ${0.02}$ &  ${0.84}$ \\
       ${--}$ & ${--}$ &  ${0.64}$ &  ${2.83}$ &  ${13.14}$ &  ${0.02}$ &  ${0.37}$ \\ \hline
        ${6369366}$ &  $${--}$$ & ${--}$ & ${--}$ &  ${12.27}$ &  ${0.03}$ &  ${0.53}$ \\
       ${--}$ & ${--}$ &  ${0.26}$ &  ${0.06}$ &  ${12.33}$ &  ${0.03}$ &  ${0.53}$ \\
       ${--}$ & ${--}$ &  ${7.56}$ &  ${3.08}$ &  ${15.35}$ &  ${0.05}$ &  ${0.20}$ \\ \hline
        ${6418986}$ &  ${2.88}$ & ${--}$ & ${--}$ &  ${10.59}$ &  ${0.02}$ &  ${0.87}$ \\
       ${--}$ & ${--}$ &  ${6.70}$ &  ${3.90}$ &  ${14.49}$ &  ${0.02}$ &  ${0.27}$ \\ \hline
        ${6468660}$ &  ${0.86}$ & ${--}$ & ${--}$ &  ${10.31}$ &  ${0.02}$ &  ${0.80}$ \\
       ${--}$ & ${--}$ &  ${2.67}$ &  ${5.27}$ &  ${15.58}$ &  ${0.03}$ &  ${0.13}$ \\ \hline
        ${6837319}$ &  ${18.85}$ & ${--}$ & ${--}$ &  ${10.96}$ &  ${0.01}$ &  ${0.90}$ \\
       ${--}$ & ${--}$ &  ${0.38}$ &  ${0.48}$ &  ${11.44}$ &  ${0.01}$ &  ${0.76}$ \\ \hline
        ${7105891}$ &  ${3.68}$ & ${--}$ & ${--}$ &  ${9.77}$ &  ${0.01}$ &  ${0.96}$ \\
       ${--}$ & ${--}$ &  ${0.55}$ &  ${1.70}$ &  ${11.47}$ &  ${0.01}$ &  ${0.56}$ \\ \hline
        ${7499271}$ &  ${9.93}$ & ${--}$ & ${--}$ &  ${9.61}$ &  ${0.02}$ &  ${0.66}$ \\
       ${--}$ & ${--}$ &  ${0.66}$ &  ${0.91}$ &  ${10.52}$ &  ${0.02}$ &  ${0.51}$ \\ \hline
        ${7499372}$ &  ${1.62}$ & ${--}$ & ${--}$ &  ${11.79}$ &  ${0.02}$ &  ${0.77}$ \\
       ${--}$ & ${--}$ &  ${2.41}$ &  ${3.16}$ &  ${14.95}$ &  ${0.03}$ &  ${0.31}$ \\
       ${--}$ & ${--}$ &  ${7.42}$ &  ${3.98}$ &  ${15.77}$ &  ${0.03}$ &  ${0.23}$ \\ \hline
        ${7799843}$ &  ${2.18}$ & ${--}$ & ${--}$ &  ${11.08}$ &  ${0.01}$ &  ${0.77}$ \\
       ${--}$ & ${--}$ &  ${0.45}$ &  ${2.80}$ &  ${13.87}$ &  ${0.01}$ &  ${0.35}$ \\ \hline
        ${7801848}$ &  ${26.56}$ & ${--}$ & ${--}$ &  ${8.60}$ &  ${0.02}$ &  ${1.12}$ \\
       ${--}$ & ${--}$ &  ${0.38}$ &  ${0.17}$ &  ${8.77}$ &  ${0.02}$ &  ${1.06}$ \\ \hline
        ${7936309}$ &  ${38.96}$ & ${--}$ & ${--}$ &  ${9.23}$ &  ${0.02}$ &  ${0.70}$ \\
       ${--}$ & ${--}$ &  ${0.19}$ &  ${0.66}$ &  ${9.89}$ &  ${0.02}$ &  ${0.57}$ \\ \hline
        ${8128926}$ &  ${1.07}$ & ${--}$ & ${--}$ &  ${11.53}$ &  ${0.05}$ &  ${0.51}$ \\
       ${--}$ & ${--}$ &  ${4.28}$ &  ${0.26}$ &  ${11.80}$ &  ${0.05}$ &  ${0.48}$ \\ \hline
        ${8263825}$ &  ${2.69}$ & ${--}$ & ${--}$ &  ${11.34}$ &  ${0.02}$ &  ${0.71}$ \\
       ${--}$ & ${--}$ &  ${0.88}$ &  ${1.98}$ &  ${13.32}$ &  ${0.02}$ &  ${0.41}$ \\ \hline
        ${8311864}$ &  ${1.01}$ & ${--}$ & ${--}$ &  ${11.87}$ &  ${0.02}$ &  ${1.05}$ \\
       ${--}$ & ${--}$ &  ${4.55}$ &  ${4.40}$ &  ${16.27}$ &  ${0.02}$ &  ${0.28}$ \\ \hline
        ${8394589}$ &  ${0.90}$ & ${--}$ & ${--}$ &  ${8.33}$ &  ${0.02}$ &  ${1.18}$ \\
       ${--}$ & ${--}$ &  ${1.98}$ &  ${2.47}$ &  ${10.80}$ &  ${0.02}$ &  ${0.53}$ \\ \hline
        ${8462696}$ &  ${3.21}$ & ${--}$ & ${--}$ &  ${10.57}$ &  ${0.02}$ &  ${0.91}$ \\
       ${--}$ & ${--}$ &  ${0.78}$ &  ${1.64}$ &  ${12.22}$ &  ${0.02}$ &  ${0.54}$ \\ \hline
        ${8800954}$ &  ${1.84}$ & ${--}$ & ${--}$ &  ${11.65}$ &  ${0.02}$ &  ${0.80}$ \\
       ${--}$ & ${--}$ &  ${1.07}$ &  ${2.40}$ &  ${14.05}$ &  ${0.02}$ &  ${0.40}$ \\ \hline
        ${9003169}$ &  ${0.86}$ & ${--}$ & ${--}$ &  ${10.10}$ &  ${0.02}$ &  ${1.13}$ \\
       ${--}$ & ${--}$ &  ${2.64}$ &  ${6.21}$ &  ${16.31}$ &  ${0.04}$ &  ${0.14}$ \\ \hline
        ${9202350}$ &  $${--}$$ & ${--}$ & ${--}$ &  ${10.24}$ &  ${0.02}$ &  ${0.68}$ \\
       ${--}$ & ${--}$ &  ${0.42}$ &  ${0.08}$ &  ${10.31}$ &  ${0.02}$ &  ${0.66}$ \\ \hline
        ${9226614}$ &  ${1.49}$ & ${--}$ & ${--}$ &  ${11.47}$ &  ${0.02}$ &  ${0.82}$ \\
       ${--}$ & ${--}$ &  ${2.87}$ &  ${4.07}$ &  ${15.54}$ &  ${0.02}$ &  ${0.24}$ \\
       ${--}$ & ${--}$ &  ${5.40}$ &  ${2.43}$ &  ${13.91}$ &  ${0.02}$ &  ${0.41}$ \\
       ${--}$ & ${--}$ &  ${7.74}$ &  ${6.27}$ &  ${17.74}$ &  ${0.05}$ &  ${0.10}$ \\ \hline
        ${9277764}$ &  ${2.37}$ & ${--}$ & ${--}$ &  ${12.17}$ &  ${0.01}$ &  ${0.78}$ \\
       ${--}$ & ${--}$ &  ${0.98}$ &  ${0.70}$ &  ${12.86}$ &  ${0.01}$ &  ${0.63}$ \\ \hline
        ${10186796}$ &  ${0.96}$ & ${--}$ & ${--}$ &  ${8.95}$ &  ${0.01}$ &  ${0.87}$ \\
       ${--}$ & ${--}$ &  ${1.49}$ &  ${2.59}$ &  ${11.54}$ &  ${0.01}$ &  ${0.41}$ \\
       ${--}$ & ${--}$ &  ${1.53}$ &  ${2.55}$ &  ${11.50}$ &  ${0.01}$ &  ${0.41}$ \\
       ${--}$ & ${--}$ &  ${6.30}$ &  ${6.70}$ &  ${15.65}$ &  ${0.05}$ &  ${0.10}$ \\ \hline
        ${10203255}$ &  ${0.98}$ & ${--}$ & ${--}$ &  ${9.98}$ &  ${0.01}$ &  ${0.81}$ \\
       ${--}$ & ${--}$ &  ${2.53}$ &  ${5.39}$ &  ${15.37}$ &  ${0.04}$ &  ${0.13}$ \\
       ${--}$ & ${--}$ &  ${4.64}$ &  ${6.66}$ &  ${16.64}$ &  ${0.08}$ &  ${0.10}$ \\ \hline
        ${10329365}$ &  ${27.87}$ & ${--}$ & ${--}$ &  ${11.55}$ &  ${0.01}$ &  ${1.38}$ \\
       ${--}$ & ${--}$ &  ${0.35}$ &  ${0.48}$ &  ${12.04}$ &  ${0.01}$ &  ${1.15}$ \\
       ${--}$ & ${--}$ &  ${4.51}$ &  ${2.98}$ &  ${14.54}$ &  ${0.02}$ &  ${0.52}$ \\ \hline
        ${10423465}$ &  ${10.83}$ & ${--}$ & ${--}$ &  ${10.24}$ &  ${0.02}$ &  ${0.86}$ \\
       ${--}$ & ${--}$ &  ${0.52}$ &  ${1.71}$ &  ${11.96}$ &  ${0.02}$ &  ${0.51}$ \\ \hline
        ${10645813}$ &  ${10.19}$ & ${--}$ & ${--}$ &  ${12.35}$ &  ${0.02}$ &  ${1.25}$ \\
       ${--}$ & ${--}$ &  ${0.85}$ &  ${0.97}$ &  ${13.33}$ &  ${0.02}$ &  ${0.88}$ \\
       ${--}$ & ${--}$ &  ${7.95}$ &  ${5.04}$ &  ${17.39}$ &  ${0.06}$ &  ${0.26}$ \\ \hline
        ${10678359}$ &  $${--}$$ & ${--}$ & ${--}$ &  ${12.01}$ &  ${0.02}$ &  ${0.51}$ \\
       ${--}$ & ${--}$ &  ${0.25}$ &  ${0.31}$ &  ${12.32}$ &  ${0.02}$ &  ${0.47}$ \\
       ${--}$ & ${--}$ &  ${0.44}$ &  ${1.59}$ &  ${13.60}$ &  ${0.02}$ &  ${0.33}$ \\ \hline
        ${10710452}$ &  ${1.19}$ & ${--}$ & ${--}$ &  ${11.07}$ &  ${0.02}$ &  ${0.80}$ \\
       ${--}$ & ${--}$ &  ${5.20}$ &  ${5.20}$ &  ${16.28}$ &  ${0.03}$ &  ${0.14}$ \\ \hline
        ${11550055}$ &  ${11.91}$ & ${--}$ & ${--}$ &  ${10.95}$ &  ${0.02}$ &  ${0.85}$ \\
       ${--}$ & ${--}$ &  ${0.39}$ &  ${1.92}$ &  ${12.87}$ &  ${0.02}$ &  ${0.48}$ \\
       ${--}$ & ${--}$ &  ${1.99}$ &  ${4.96}$ &  ${15.91}$ &  ${0.03}$ &  ${0.17}$ \\ \hline
        ${11714337}$ &  ${1.61}$ & ${--}$ & ${--}$ &  ${12.24}$ &  ${0.02}$ &  ${0.86}$ \\
       ${--}$ & ${--}$ &  ${0.85}$ &  ${1.04}$ &  ${13.28}$ &  ${0.02}$ &  ${0.62}$ \\
       ${--}$ & ${--}$ &  ${5.17}$ &  ${4.54}$ &  ${16.78}$ &  ${0.04}$ &  ${0.21}$ \\ \hline
        ${12068968}$ &  ${21.72}$ & ${--}$ & ${--}$ &  ${11.44}$ &  ${0.02}$ &  ${1.05}$ \\
       ${--}$ & ${--}$ &  ${0.31}$ &  ${0.78}$ &  ${12.21}$ &  ${0.02}$ &  ${0.80}$ \\ \hline
    \caption{Companion Properties}
    \label{tab:comps}
\end{longtable*}

\section{Radius Correction Factor Distributions}\label{app:corr}

Tables~\ref{tab:X1} and \ref{tab:X2} present the distributions of planet radius correction factors  used in our study assuming planets orbit the primary or secondary star, respectively. Details on how these correction factors were determined are included in Section~\ref{sec:rcorr}, and their distributions are visualized in Figure~\ref{fig:corrections}. Each table includes the binning details for each correction factor distribution, the number (and relative probability) of observed and simulated undetected companions whose correction factors fall within that bin, and the average probability (between the observed and simulated populations) of a star having a correction factor in that bin.

\begin{table}[!htpb]
    \centering
    \begin{tabular}{|cc|cc|cc|c|}
    \hline
 \multicolumn{2}{|c}{Bin Edges [R$_\odot$]}& \multicolumn{2}{|c|}{Observed}& \multicolumn{2}{c|}{Simulated Undetected}&Average Probability [$\%$]\\
 Lower& Upper& Counts& Probability [$\%$]& Counts& Probability [$\%$]&\\ \hline\hline
        ${1.00}$ & ${1.04}$ & ${20}$ & ${50.00}$ & ${248842}$ & ${54.16}$ & ${52.08}$ \\
        ${1.04}$ & ${1.08}$ & ${2}$ & ${5.00}$ & ${65811}$ & ${14.32}$ & ${9.66}$ \\
        ${1.08}$ & ${1.12}$ & ${3}$ & ${7.50}$ & ${37683}$ & ${8.20}$ & ${7.85}$ \\
        ${1.12}$ & ${1.17}$ & ${4}$ & ${10.00}$ & ${25880}$ & ${5.63}$ & ${7.82}$ \\
        ${1.17}$ & ${1.21}$ & ${2}$ & ${5.00}$ & ${19218}$ & ${4.18}$ & ${4.59}$ \\
        ${1.21}$ & ${1.25}$ & ${1}$ & ${2.50}$ & ${14741}$ & ${3.21}$ & ${2.85}$ \\
        ${1.25}$ & ${1.29}$ & ${2}$ & ${5.00}$ & ${12124}$ & ${2.64}$ & ${3.82}$ \\
        ${1.29}$ & ${1.33}$ & ${1}$ & ${2.50}$ & ${12144}$ & ${2.64}$ & ${2.57}$ \\
        ${1.33}$ & ${1.37}$ & ${4}$ & ${10.00}$ & ${14336}$ & ${3.12}$ & ${6.56}$ \\
        ${1.37}$ & ${1.41}$ & ${1}$ & ${2.50}$ & ${8657}$ & ${1.88}$ & ${2.19}$ \\
        \hline
    \end{tabular}
    \caption{Average Probability for Radius Correction Factors Assuming Planet Orbits the Primary (X$_1$). Results are divided across 10 bins of equal width spanning from 1 to $\sqrt{2}$.}
    \label{tab:X1}
\end{table}

\begin{table}[!htpb]
    \centering
    \begin{tabular}{|cc|cc|cc|c|}
    \hline
\multicolumn{2}{|c}{Bin Edges [R$_\odot$]}& \multicolumn{2}{|c}{Observed}& \multicolumn{2}{|c|}{Simulated Undetected}&Average Probability [$\%$]\\
 Lower& Upper& Counts& Probability [$\%$]& Counts& Probability [$\%$]&\\ \hline\hline
        ${1.00}$ & ${1.65}$ & ${15}$ & ${37.50}$ & ${140002}$ & ${30.88}$ & ${34.19}$ \\
        ${1.65}$ & ${2.30}$ & ${6}$ & ${15.00}$ & ${158508}$ & ${34.96}$ & ${24.98}$ \\
        ${2.30}$ & ${2.95}$ & ${6}$ & ${15.00}$ & ${96798}$ & ${21.35}$ & ${18.18}$ \\
        ${2.95}$ & ${3.60}$ & ${5}$ & ${12.50}$ & ${48268}$ & ${10.65}$ & ${11.57}$ \\
        ${3.60}$ & ${4.25}$ & ${1}$ & ${2.50}$ & ${8960}$ & ${1.98}$ & ${2.24}$ \\
        ${4.25}$ & ${4.90}$ & ${5}$ & ${12.50}$ & ${643}$ & ${0.14}$ & ${6.32}$ \\
        ${4.90}$ & ${5.55}$ & ${0}$ & ${0.00}$ & ${122}$ & ${0.03}$ & ${0.01}$ \\
        ${5.55}$ & ${6.20}$ & ${1}$ & ${2.50}$ & ${43}$ & ${0.01}$ & ${1.25}$ \\
        ${6.20}$ & ${6.85}$ & ${0}$ & ${0.00}$ & ${0}$ & ${0.00}$ & ${0.00}$ \\
        ${6.85}$ & ${7.50}$ & ${1}$ & ${2.50}$ & ${0}$ & ${0.00}$ & ${1.25}$ \\
        \hline
    \end{tabular}
    \caption{Average Probability for Radius Correction Factors Assuming Planet Orbits the Secondary (X$_2$). Results are divided across 10 bins of equal width spanning from 1 to 7.5.}
    \label{tab:X2}
\end{table}

\section{Updating \kepler{} Survey Completeness from Bergsten et al. (2022)}\label{app:2022updates}

\citet{Bergsten2022} made use of sample-specific average completeness maps computed following the methodology of \citet{Mulders2018}. This involved calculating an average detection efficiency map using \texttt{KeplerPORTS} outputs for all stars in the sample, and a geometric transit probability map using the sample's average stellar properties.

In this work, we made use of per-star completeness maps where the detection efficiency and geometric transit probability were evaluated independently for each star\footnote{As in \citet{Bergsten2023}, but focusing on FGK stars instead of M dwarfs.}. Evaluating \kepler{}'s survey completeness at the per-star level gave us the control necessary to directly modify completeness maps for individual stars when simulating the effects of stellar companions, described Section~\ref{sec:occ_sim}. We note that the per-star maps are still averaged together when optimizing occurrence models, but the use of per-star maps from the outset allows for greater capture of per-star sensitivity variations and elements of completeness (like the geometric transit probability) that depend on stellar properties. 

\section{Best-Fit Model Parameters}\label{app:params}

Table~\ref{tab:params} presents the best-fit occurrence model free parameters -- using the parametric model defined in \citet{Bergsten2022} -- for the variety of scenarios examined in this work. Each parameter includes a median estimate based on the posterior distribution of that scenario's Markov Chain Monte Carlo fits, along with \nth{16} and \nth{84} percentile values from those distributions to represent lower and upper $1\sigma$ uncertainties, respectively. The occurrence model and optimization process are described in Section~\ref{sec:occ}, and the median occurrence distributions are presented in Figure~\ref{fig:occ_marg}.

\begin{sidewaystable}
    \centering
    \begin{tabular}{|lc|cccccccc|}
    \hline
          &Scenario&  $F_\mathrm{0}$&  $P_\mathrm{break}$&  $\beta_1$&  $\beta_2$&  $P_\mathrm{central}$&  $s$&  $\chi_1$& $\chi_2$\\
  && & [days]& & & [days]& [days]& &\\ \hline\hline
         &No treatment of stellar companions& ${0.67}^{+0.02}_{-0.02}$ & ${10.28}^{+1.24}_{-1.75}$ & ${0.55}^{+0.18}_{-0.11}$ & ${-0.92}^{+0.09}_{-0.08}$ & ${11.84}^{+1.77}_{-1.56}$ & ${2.62}^{+0.42}_{-0.37}$ & ${0.83}^{+0.02}_{-0.02}$ & ${0.35}^{+0.04}_{-0.05}$ \\ \hline
         \multirow{3}{*}{Observed Rate (20\%)}&Planets orbit primary& ${0.69}^{+0.02}_{-0.02}$ & ${10.02}^{+1.37}_{-1.84}$ & ${0.58}^{+0.21}_{-0.12}$ & ${-0.91}^{+0.10}_{-0.09}$ & ${11.85}^{+1.77}_{-1.57}$ & ${2.62}^{+0.43}_{-0.38}$ & ${0.84}^{+0.02}_{-0.02}$ & ${0.36}^{+0.04}_{-0.05}$ \\
         &Planets orbit secondary& ${0.78}^{+0.03}_{-0.02}$ & ${9.90}^{+1.41}_{-1.85}$ & ${0.59}^{+0.21}_{-0.13}$ & ${-0.89}^{+0.09}_{-0.09}$ & ${11.94}^{+1.77}_{-1.57}$ & ${2.64}^{+0.42}_{-0.38}$ & ${0.85}^{+0.02}_{-0.02}$ & ${0.36}^{+0.04}_{-0.05}$ \\
         &Mix of primary/secondary orbits& ${0.73}^{+0.02}_{-0.02}$ & ${9.98}^{+1.39}_{-1.86}$ & ${0.58}^{+0.21}_{-0.12}$ & ${-0.90}^{+0.10}_{-0.09}$ & ${11.89}^{+1.78}_{-1.58}$ & ${2.63}^{+0.42}_{-0.38}$ & ${0.84}^{+0.02}_{-0.02}$ & ${0.36}^{+0.04}_{-0.05}$ \\ \hline
         \multirow{3}{*}{Field Rate (45\%)}&Planets orbit primary& ${0.70}^{+0.02}_{-0.02}$ & ${9.93}^{+1.41}_{-1.85}$ & ${0.59}^{+0.21}_{-0.13}$ & ${-0.90}^{+0.10}_{-0.09}$ & ${11.85}^{+1.81}_{-1.60}$ & ${2.62}^{+0.43}_{-0.38}$ & ${0.84}^{+0.02}_{-0.02}$ & ${0.37}^{+0.04}_{-0.05}$ \\
         &Planets orbit secondary& ${0.96}^{+0.03}_{-0.03}$ & ${9.17}^{+1.75}_{-1.59}$ & ${0.66}^{+0.22}_{-0.16}$ & ${-0.85}^{+0.09}_{-0.10}$ & ${12.23}^{+1.69}_{-1.59}$ & ${2.69}^{+0.41}_{-0.39}$ & ${0.87}^{+0.02}_{-0.02}$ & ${0.39}^{+0.04}_{-0.05}$ \\
         &Mix of primary/secondary orbits& ${0.81}^{+0.03}_{-0.03}$ & ${9.64}^{+1.53}_{-1.79}$ & ${0.62}^{+0.22}_{-0.14}$ & ${-0.88}^{+0.09}_{-0.09}$ & ${12.01}^{+1.77}_{-1.59}$ & ${2.65}^{+0.43}_{-0.39}$ & ${0.85}^{+0.02}_{-0.02}$ & ${0.38}^{+0.04}_{-0.05}$ \\
    \hline
    \end{tabular}
    
    \caption{Best-fit parameters for the \citet{Bergsten2022} model for each scenario and treatment in this study.}
    \label{tab:params}
\end{sidewaystable}

\end{document}